\documentclass[aip,pop,graphicx,amsmath,amssymb,preprint,numerical]{revtex4-1}

\usepackage{graphicx}
\usepackage{dcolumn}
\usepackage{bm}

\usepackage[utf8]{inputenc}
\usepackage[T1]{fontenc}
\usepackage{mathptmx}
\usepackage{etoolbox}

\graphicspath{{./}{figures/}}
\usepackage{epstopdf}
\newcommand\wpe{\omega_\mathrm{pe}}
\newcommand\wce{\Omega_\mathrm{ce}}
\newcommand\lde{\lambda_\mathrm{De}}
\newcommand\degr{\circ}
\makeatletter
\def\@email#1#2{%
 \endgroup
 \patchcmd{\titleblock@produce}
  {\frontmatter@RRAPformat}
  {\frontmatter@RRAPformat{\produce@RRAP{*#1\href{mailto:#2}{#2}}}\frontmatter@RRAPformat}
  {}{}
}%
\makeatother
\begin{document}


\title{Plasma emission induced by ring-distributed energetic electrons in overdense plasmas}

\author{Yao Chen}
\email{yaochen@sdu.edu.cn}
\affiliation{Institute of Frontier and Interdisciplinary Science, Shandong University, Qingdao, Shandong, 266237, People's Republic of China.}
\affiliation{Institute of Space Sciences, Shandong University, Shandong, 264209, People's Republic of China.}

\author{Zilong Zhang}
\affiliation{Institute of Space Sciences, Shandong University, Shandong, 264209, People's Republic of China.}
\affiliation{Institute of Frontier and Interdisciplinary Science, Shandong University, Qingdao, Shandong, 266237, People's Republic of China.}

\author{Sulan Ni}
\affiliation{Institute of Frontier and Interdisciplinary Science, Shandong University, Qingdao, Shandong, 266237, People's Republic of China.}
\affiliation{Institute of Space Sciences, Shandong University, Shandong, 264209, People's Republic of China.}

\author{Hao Ning}
\affiliation{Institute of Frontier and Interdisciplinary Science, Shandong University, Qingdao, Shandong, 266237, People's Republic of China.}
\affiliation{Institute of Space Sciences, Shandong University, Shandong, 264209, People's Republic of China.}

\author{Chuanyang Li}
\affiliation{Institute of Frontier and Interdisciplinary Science, Shandong University, Qingdao, Shandong, 266237, People's Republic of China.}
\affiliation{Institute of Space Sciences, Shandong University, Shandong, 264209, People's Republic of China.}

\author{Yaokun Li}
\affiliation{Institute of Frontier and Interdisciplinary Science, Shandong University, Qingdao, Shandong, 266237, People's Republic of China.}
\affiliation{Institute of Space Sciences, Shandong University, Shandong, 264209, People's Republic of China.}

\date{\today}

\begin{abstract}
According to the standard scenario of plasma emission, escaping radiations are generated by the nonlinear development of the kinetic bump-on-tail instability driven by a single beam of energetic electrons interacting with plasmas. Here we conduct fully-kinetic electromagnetic particle-in-cell simulations to investigate plasma emission induced by the ring-distributed energetic electrons interacting with overdense plasmas. Efficient excitations of the fundamental (F) and harmonic (H) emissions are revealed with radiation mechanism(s) different from the standard scenario: (1) The primary modes accounting for the radiations are generated through the electron cyclotron maser instability (for the upper-hybrid (UH) and Z modes) and the thermal anisotropic instability (for the whistler (W) mode); the F emission is generated by the nonlinear coupling of the Z and W modes and the H emission by the nonlinear coupling of the UH modes. This presents an alternative mechanism of coherent radiation in overdense plasmas.
\end{abstract}

\maketitle

\section{Introduction}

Plasma emission (PE) refers to coherent radiation from plasmas at frequencies
close to the fundamental plasma frequency ($\wpe$) and/or its harmonics.
\citet{1958SvA.....2..653G} proposed the first theoretical framework
of PE to explain the highly non-thermal solar radio bursts
discovered in the 1940-1950s (c.f., Refs.~\onlinecite{1950AuSRA...3..399W,1950AuSRA...3..541W,1954AuJPh...7..439W,1985srph.book.....M}
). The original framwork has been refined over
the past decades (e.g., Refs.~\onlinecite{1970AuJPh..23..871M,1979SoPh...61..161T,1985JGR....90.6637C,1987JPlPh..38..169C,1988JGR....93.3958C,1993ApJ...408..720R,1994ApJ...422..870R}, reviewed by Refs.~\onlinecite{1980SSRv...26....3M,1986islp.book.....M,2017RvMPP...1....5M}).
The refined model is referred to as the standard paradigm of PE hereafter.

According to the standard paradigm (and the above references), the radiating
process starts from the injection of a beam of energetic electrons
into background plasmas which drives the kinetic bump-on-tail
instability and generates electrostatic Langmuir (L) waves through
couplings with the beam mode. The fundamental (F) and harmonic (H)
electromagnetic radiations are then generated through subsequent
nonlinear wave-wave coupling processes including (1) the
electrostatic decay of the primary L wave to generate the
secondary backward-propagating L wave and the ion acoustic (IA) mode
in terms of L $\rightarrow$ L$^\prime$ + IA; (2) the electromagnetic decay of the primary
L wave to generate the F radiation and IA mode (L $\rightarrow$ F + IA);
(3) the resonant coupling of the L wave and the IA wave to generate
the F emission (L + IA $\rightarrow$ F), as an alternative mechanism of the
F emission; the above two processes can be put together as
(L $\pm$ IA $\rightarrow$ F); (4) the resonant coupling of the forward-propagating L wave
and the backward-propagating L$^\prime$ wave to generate the H emission (L + L$^\prime$ $\rightarrow$ H).

Other scenarios of PE have been proposed, such as
(1) the linear mode conversion to release the F emission with the L
wave propagating in inhomogeneous plasmas, within such plasmas
energy can transfer from one mode to the other since they are
coupled in the frequency-wave vector space
(e.g., Refs.~\onlinecite{1985JGR....90.6637C,1992PhFlB...4.1772H,2007PhRvL..99a5003K}); (2) the quasi-mode mechanism for the
H emission from the Earth's bow shock, according to which the nonlinear
interaction between the electron beam and the backscattered L wave generates
the quasi-mode with frequency at 2 $\wpe$ which can
convert to the H emission when propagating in inhomogeneous plasmas
with slightly decreasing density \cite{1994JGR....9923481Y}; (3) the antenna
radiations at both the F and H frequencies by the localized Langmuir
currents as eigenmodes of solar wind density cavities, this mechanism
does not require the presence of backscattered L wave to generate the
H emission \cite{1978GeoRL...5..881P,1980PhFl...23..388G,2010JGRA..115.1101M,2011GeoRL..3813101M,2012ApJ...755...45M,2013JGRA..118.6880M}.

Another group of PE involves the interaction of the L mode with the W mode,
first suggested by \citet{1970SoPh...13..420C} and  \citet{1972P&SS...20..711C}. The idea was rejected by
\citet{1975AuJPh..28..101M} who concluded that the only wave around the plasma
frequency that could coalesce with whistlers (to give the F emission) is
the superluminal Z mode with a small wave number. Yet, \citet{1975AuJPh..28..101M} further
deduced that the Z + W $\rightarrow$ O/F process (`O' for the ordinary electromagnetic mode and `W' for the whistler mode)
can occur only under very restrictive condition and therefore PE involving
whistlers is unlikely to be important, at least not in coronal plasmas. This
deduction was based on the evaluation of the matching conditions of three-wave
interaction using over-simplified disperion relations. After solving the
complete magnetoionic dispersion relations, \citet{2021PhPl...28d0701N}
concluded that the matching conditions of the three-wave interaction (Z + W$\rightarrow$ O/F)
can be satisfied over a wide regime of parameters. They also demonstrated
the occurrence of such process with wave-pumping PIC simulations.

It is essential to verify the proposed nonlinear multi-stage PE process
with self-consistent fully-kinetic electromagnetic and relativistic
simulation with the least approximation to basic laws of mechanics and
electromagnetism. Indeed, researches along this line have become one major
front of studies on PE (e.g., Refs.~\onlinecite{2001JGR...10618693K,2009ApJ...694..618R,2010JGRA..115.1204U,2011PhPl...18e2903T,2012SoPh..280..551G,2015A&A...584A..83T,2019JGRA..124.1475H,2019ApJ...871...74L}). In addition
to studies on PE in homogeneous plasmas, effect of density fluctuations
of solar wind plasmas on the properties of both the F and H emissions has been
investigated with PIC simulations \cite{2021ApJ...923..103K,2022ApJ...924L..24K}.

A majority of these PIC simulations of PE were designed
for the beam distribution of energetic electrons, agreeing
with the assumption of the standard paradigm of PE.
Yet, non-beam distribution exists pervasively in space
and astrophysical plasmas. The shape of velocity distribution
functions (VDFs) depends on the details of particle acceleration
through either shock (e.g., Refs.~\onlinecite{1989GeoRL..16.1125W,
1984AnGeo...2..449L, 1987ApJ...322..463V, 1987SoPh..111..155V,
2020ApJ...900L..24Y}) or magnetic reconnection (e.g., Refs.~\onlinecite{2014GeoRL..41.8688B,
2014GeoRL..41.5389S,2015GeoRL..42.2586S}), and their interaction
with waves/fluctuations and inhomogeneous magnetic fields and plasmas.
For instance, beam electrons can easily transform into ring-beam or
ring-type distributions if being injected into field lines not
along the beam direction, or into the loss-cone type distribution
by the mirror effect if propagating into inhomogeneous magnetic field.
These distributions can energize various wave modes whose couplings
may lead to coherent radiation.

Thus, the paradigm of coherent plasma emission (in overdense or weakly-magnetized plasmas with large ratio of $\wpe$ and $\wce$ (the electron gyro-frequency))
should be further developed to address whether and how radiations
generated in such plasmas interacting with these non-beam electrons. This is crucial to
understanding solar radio bursts (and other radio bursts in space) that
may stem from non-beam energetic electrons, such as the type-I
noise storm and its continuum (e.g., Ref.~\onlinecite{2017SoPh..292...82L}),
and type-II, IV, and V bursts (see, e.g., Refs.~\onlinecite{2012ApJ...753...21F,
2012ApJ...750..158K, 2014ApJ...787...59C, 2016ApJ...830L...2V, 2019ApJ...870...30V}).

As one such example, \citet{2020ApJ...891L..25N} performed PIC simulations to
investigate the radiation process due to the interaction of
energetic electrons of the Dory--Guest--Harris
(DGH \cite{1965PhRvL..14..131D}) distribution --- a kind of double-sided
loss-cone distribution --- with overdense plasmas. They deduced that (1) the primary mode is the upper hybrid (UH) mode, weaker modes in terms of Z and W also
present; these modes are excited via the electron cyclotron maser
instability (ECMI), (2) the F and H emissions can be generated via a
multi-stage nonlinear process that is similar yet different from the
standard paradigm, with the nonlinear coupling of almost
counter-propagating UH modes leading to the H emission
(referred to as `UH + UH$^\prime\ \rightarrow$ H') and
that of the almost counter-propagating Z and W modes
leading to the F emission in the O mode (Z + W $\rightarrow$ O/F).
This presents an alternative possibility of PE in overdense plasmas.

PIC simulations with other type of distribution of energetic electrons
are required to further verify the above scenario of PE.
One natural candidate is the ring-beam distribution which may stem from
the beam distribution as mentioned. Previous PIC studies of coherent
radiation with ring-beam electrons were mainly for underdense or strongly-magnetized plasmas with a relatively small $\wpe/\wce~(< 1)$ where $\wce$ represents the electron cyclotron frequency, in which another kind
of coherent radiation in plasmas, i.e., the electron cyclotron maser
emission (ECME: Ref.~\onlinecite{1979ApJ...230..621W}), plays a major role
(e.g., Refs.~\onlinecite{2009PhRvL.103j5101L,2011PhPl...18i2110L, 2020ApJ...891...92Z}).
Studies for overdense plasmas with a large $\wpe/\wce~(> 1)$ have
failed to obtain significant F/H plasma emissions (e.g., Ref.~\onlinecite{2020ApJ...891...92Z}),
likely due to the following three factors: (1) according
to \citet{2020ApJ...891...92Z}, the spectral resolution in $\omega$ of their
simulation is about 0.015 $\wpe$ that is not small enough to distinguish
the O and X modes and resolve the escaping radiation; (2)
the simulation domain size is 1024 $\lde$,
much shorter than the expected wavelength of the F emission,
thus neither the wave number resolution is enough to
resolve the expected O/F mode (see, e.g.,
Ref.~\onlinecite{2022ApJ...924L..34C}); (3) a large abundance (varying from 5{\%} to 50{\%})
of ring-beam electrons is employed, this may cause the mismatch of
the resonance conditions of three-wave
interaction (L + IA $\rightarrow$ O/F) since the frequency of the beam-driven Langmuir wave
may be significantly below $\wpe$ due to its coupling with the beam mode.
(c.f., Refs.~\onlinecite{1989PhFlB...1..204C,2015A&A...584A..83T}).

To simplify the problem, it is necessary to distinguish the effect
of the ring component from the beam to clarify the underlying
physics of radiations. This is done here by simulating the
effect of energetic electrons with a pure-ring
distribution and comparing with latest published results for
pure-beam distribution such as \citet{2019JGRA..124.1475H},
\citet{2022ApJ...924L..34C} and \citet{2022arXiv220911707Z}. The next section presents the numerical
setup, followed by the analysis and comparison of three cases.
Conclusions and discussion are provided in the last section.

\section{Simulation Code and Setup}

Following \citet{2022ApJ...924L..34C} we perform the fully-kinetic electromagnetic simulation with the open-source Vector PIC \citep[VPIC:][]{2008PhPl...15e5703B,bowers20080,2009JPhCS.180a2055B} code released by the Los Alamos National Labs, run in two spatial dimensions ($x$, $z$) with three velocity and field components. The background magnetic field is along the z direction ($\vec B_0 = B_0 \hat{z}$), and the wave vector ($\vec k$) is in the $x$O$z$ plane. Periodic boundary conditions are used. The simulation domain is taken to be 5000 $\times$ 5000 $\lambda_\mathrm{De}^2$ where $\lambda_\mathrm{De}$ is the electron Debye length, the cell size $\Delta = 2.44~\lambda_\mathrm{De}$, the time step $\delta_t = 0.0217~\wpe^{-1}$, the total simulation time is 3000 $\wpe^{-1}$. The obtained frequency resolution ($\delta \omega$) is $\sim$ 0.002 $\wpe$ with Fourier analysis over data duration of 2700 $\wpe^{-1}$, and the wavenumber resolution ($\delta k$) is $\sim$ 0.7 $\wce/c$ ($\sim$ 0.0012 $\lde^{-1}$). The number of macro-particles per cell per species is taken to be 2000 for energetic electrons and 1000 for both background electrons and protons, in total 1.68-billion macro-particles are included. The density ratio of energetic electrons and the background electrons ($n_\mathrm{e}/n_\mathrm{0})$ is set to be 0.01, the temperature of the background electrons and protons ($T_\mathrm{p} = T_\mathrm{e})$ is set to be 2 MK, and the ratio of characteristic frequencies ($\wpe/\wce$) is set to be 10.

The general ring-beam distribution is expressed as
\begin{equation}
f_\mathrm{e} = A_1 \exp \left[-\left(\frac{u_{\perp}-u_\mathrm{d \perp}}{\sqrt{2}  u_{\perp0}}\right)^{2}-\left(\frac{u_{\parallel}-u_\mathrm{d \parallel}}{\sqrt{2} u_{\parallel0}}\right)^{2}\right]
\end{equation}
where $A_1$ is the normalization factor, $u_\perp (u_\parallel)$ is the perpendicular (parallel) momentum per mass, $  u_{\perp0} ( u_{\parallel0})$ is the corresponding thermal velocity component of energetic electrons, and $u_\mathrm{d\perp} (u_\mathrm{d\parallel})$ is the corresponding velocity component at the center of the ring-beam distribution. Linear instabilities driven by energetic electrons with this type of ring-beam distribution have been analyzed earlier by, e.g., \citet{2012PhPl...19g2107U} and \citet{2015PhPl...22b2112H}. The parallel and the perpendicular widths of the ring-beam Gaussian ($u_{\perp0}$ and $u_{\parallel0}$) are set to be $0.018 c$. The pure-ring distribution corresponds to $u_\mathrm{d\parallel} = 0$ or $\alpha = 90^{\degr}$, and the pure-beam distribution corresponds to $u_\mathrm{d\perp} = 0$ or $\alpha = 0^{\degr}$, where $\alpha~(= \arctan { (u_\mathrm{d\perp}/u_\mathrm{d\parallel})}$) represents the average pitch angle of the ring-beam electrons. The present study investigates the pure-ring distribution, therefore $u_\mathrm{d \parallel}$ is set to be 0. Two cases with the realistic mass ratio ($m_\mathrm{p}/m_\mathrm{e} $ = 1836) are simulated with Case A for $u_\mathrm{d\perp} =$ 0.3c and Case B for $u_\mathrm{d\perp} =$ 0.2c. Another case (Case C) for $u_\mathrm{d\perp} =$ 0.3c with a much-larger mass ratio ($m_\mathrm{p}/m_\mathrm{e} $ = 183600) is also presented since according to \citet{2022ApJ...924L..34C} the mass ratio can affect the characteristics of the beam-Langmuir and ion-acoustic (IA) modes, useful to clarify their role in the plasma emission process. Other configurations are the same for all cases presented. The variation of the total energy of the numerical system remains below 2$\times 10^{-5} E_0$ during the simulation, where $E_0$ represents the total initial energy of all electrons.

Two additional cases are also simulated, one with $u_\mathrm{d\perp} =$ 0.1c and the realistic mass ratio, and the other is the thermal case without any energetic electrons. It was found that the growth of all relevant modes is insignificant in the $ u_\mathrm{d\perp} =$ 0.1c case, in other words, their levels are close to the thermal situation. We confirm that all wave modes presented here in Cases A--C are much stronger than those of the thermal case, meaning they are indeed excited by the ring electrons rather than being some numerical noise.

\section{Wave Modes and Plasma Emissions Excited by the Ring-distributed Energetic Electrons}

We first exhibit the wave modes excited in Case A and investigate their generation mechanism mainly by analyzing their characteristics and the matching conditions of wave-particle and wave-wave resonant couplings, then we show Cases B and C to reveal more clues about the underlying physics.

\subsection{Case A with $u_\mathrm{d\perp} = 0.3c$}

In the upper panels of Figure 1 we present the energy curves of the six field components and the negative change of the total electron kinetic energy ($-\Delta E_\mathrm{k}$), in unit of $E_\mathrm{k0}$(A) which represents the total initial kinetic energy of the ring electrons in this case. It is different for different cases. For this case $E_\mathrm{k0}$(A)$~= 3.214~m_\mathrm{e0}c^2$. The kinetic evolution of the ring-plasma system can be separated into three stages, 0--300 $\wpe^{-1}$ for Stage I, 300--1000 $\wpe^{-1}$ for Stage II, and $>1000$ $\wpe^{-1}$for Stage III. Stage I is characterized by the rapid rise of $E_x$, $E_z$, and $B_z$, they reach the maximum energy at the end of this stage. In Stage II, they damp gradually. Stage III is characterized by their further damping and the persistent rise of $B_x$ and $B_y$, which get stronger than $E_x$ and $E_z$ eventually.

We select four representative moments to plot the electron VDFs in the upper panels of Figure 2 (Multimedia view). The electrons diffuse rapidly towards both larger and smaller $v_\perp$ during Stage I. Then the VDF of energetic electrons expands gradually and occupies a large velocity-space square extending from $v_\perp \sim 0.1c$ to $v_\perp \sim 0.45c$. During Stage III, the VDF manifests a pair of diffusion signatures around the lines of $v_\parallel = \pm 0.1c$, one of which is along the resonance curve of the $W_H$ mode plotted in Figure 2d, indicating strong coupling between the energetic electrons and the $W_H$ mode. The background VDF does not evolve much.

The wave modes excited by the ring electrons can be revealed with the three-dimensional (3D) Fourier analysis. Figure 3 (Multimedia view) presents the obtained wave distribution with the maximum intensity (among all relevant frequencies) in the wave-vector ($\vec k$) space, Figure 4 (Multimedia view) presents the obtained dispersion relations for the selected propagation angle ($\theta$) and range of $\omega$--$k$. The analytical dispersion curves of the four magnetoionic modes (X, O, Z, and W) are superposed. The temporal evolution of the $\vec k$-space wave map is presented in the movie accompanying Figure 3, and the complete dispersion diagrams from $\theta = 0^{\degr}$ to $\theta = 90^{\degr}$ are presented in the movie accompanying Figure 4. In the lower panels of Figure 1, we have plotted the temporal energy profiles for various modes. These figures and movies should be combined to tell the nature and characteristics of each mode.

The ring-plasma system generates two primary modes, one is the electrostatic upper-hybrid (UH) mode, the other is the electromagnetic W mode; together with the four less-intensive modes including the Z mode \citep[also referred to as the generalized Langmuir mode, see][]{2022ApJ...924L..34C}, the IA mode, and the two escaping radiation modes (i.e., the O/F and H emissions). Their characteristics and generation mechanisms will be analyzed in the following text.

\subsubsection{The UH and IA modes.}

 According to Figures 3 {\&} 4 (and the accompanying movies), the UH mode has two components, one propagates perpendicularly with $\theta \approx 90^{\degr}$ (referred to as UH1), the other propagates quasi-perpendicularly (UH2). Both have almost the same $|k_\perp|$ ($\approx 50~\wce/c$, referred to as $|k_\perp^\mathrm{UH}|$) yet UH1 is much stronger than UH2. The two components appear in the $\vec k$-space map (see the movie accompanying Figure 3) almost simultaneously, UH2 dissipates after $\sim$ 1000 $\wpe^{-1}$ while UH1 maintains a strong level of intensity during the simulation. To address their generation mechanism, in Figure 2b we plot the resonance curves with the selected parameters (see arrows in Figure 3 and parameters listed in Table 1) according to the following resonance condition of the electron cyclotron maser instability (ECMI)
\begin{equation}
\gamma \omega - k_\parallel u_\parallel - n \Omega_\mathrm{ce0}=0
\end{equation}
where $n$ is the harmonic number, $\Omega_\mathrm{ce0}$ is the gyro-frequency for electrons at rest, and $\gamma$ is the Lorentz factor. Here the strongest excitation of the UH1 mode is along the perpendicular direction with $k_\parallel = 0$. Therefore, to plot its resonance curve we set the harmonic number $n$ to be 10. It is clear that the UH1 component is excited by the ECMI since the corresponding resonance curve crosses the VDF region with significant positive gradient, while the UH2 curve crosses the negative-gradient region of the background VDF therefore it is not driven by the ECMI and its later damping is due to thermal absorption by the background plasmas.

The IA mode is very weak in energy (see Figure 4f), likely due to the strong Landau damping in plasmas with equal $T_\mathrm{p}$ and $T_\mathrm{e}$. To further examine its characteristics, we show the dispersion relations of the density fluctuations of electrons ($\delta n_\mathrm{e}~(= n_\mathrm{e} - n_\mathrm{0}) $) and protons ($\delta n_\mathrm{p}~(=n_\mathrm{p} - n_\mathrm{0})$) in Figure 5 (Multimedia view for the temporal evolution). The UH mode appears only in the $\delta n_\mathrm{e}$ spectra due to its relatively high frequency ($\sim \wpe$) and electrostatic nature. Except this difference, the other fluctuations are identical meaning they are the charge-neutral low-frequency IA mode.

The IA mode has two components, one with a large $\lvert k_\perp \rvert$ ($\approx 100~\wce/c \sim 2~|k_{\perp}^\mathrm{UH}|$, referred to as IA1), one with a nearly zero $|k_\perp|$ (IA2), i.e., parallel-propagating. Both components have comparable intensities, and both have frequencies not larger than 0.1 $\wce$ with almost the same range of $k_\parallel$ as the UH mode, despite their large difference in $k_\perp$. According to the movie accompanying Figure 5, the IA1 mode emerges after 200 $\wpe^{-1}$ following the appearance of the UH mode, and IA2 emerges after 500 $\wpe^{-1}$. On the basis of these analyses, we suggest that both IA1 and IA2 originate from the nonlinear decay of the dominant UH mode (UH1 or UH2), in terms of UH $\rightarrow$ IA + UH$^\prime$ where UH$^\prime$ represents the daughter UH mode propagating either along or opposite to its mother UH mode. According to the above obtained ranges of $(\omega, \vec k)$, the corresponding matching conditions can be satisfied.

\subsubsection{The W mode.}

 The W mode is one of the two primary modes, with magnetic field components ($B_x$, $B_y$) much stronger than its electric field components (by $\sim 2$ orders in magnitude). Its total energy can reach above $10^{-3}~E_\mathrm{k0}$(A) according to Figure 1. It also consists two components, one high-frequency component ($\sim 0.4~\wce$, referred to as W$_\mathrm{H}$) and one low-frequency component ($\le 0.1~ \wce$, W$_\mathrm{L}$), with W$_\mathrm{H}$ being much stronger than W$_\mathrm{L}$.  W$_\mathrm{H}$ propagates mainly along parallel and quasi-parallel directions ($|\theta|<20^{\degr}$), and W$_\mathrm{L}$ presents a quadrupolar pattern of propagation (see Figures 3 {\&} 4). In addition, according to the movie accompanying Figure 3 W$_\mathrm{L}$ reaches its maximum intensity before 700 $\wpe^{-1}$ while W$_\mathrm{H}$ becomes stronger than W$_\mathrm{L}$ after 1000 $\wpe^{-1}$ and increases hereafter persistently in intensity. These discrepancies indicate their different physical origin. To clarify this, we plot the resonance curves (see Figure 2d) with parameters of ($\omega, \vec k$) listed in Table 1 (also see arrows in Figure 4c--d). The curve of W$_\mathrm{H}$ crosses significant positive gradient of VDF while is \emph{not} for W$_\mathrm{L}$ whose resonance curve is too large to be visible in Figure 2 for the adopted ranges of coordinates. This means that the W$_\mathrm{H}$ is excited by the ECMI and W$_\mathrm{L}$ \emph{NOT}.

The thermal anisotropy if strong enough can excite the low-frequency W mode. The following equation gives the threshold condition of this instability \cite{1983bpp..conf..229D}
\begin{equation}
\frac{T_\mathrm{e\perp}}{T_\mathrm{e\parallel}} > 1 + \frac{c^2 k^2}{\wpe^2}
\end{equation}
For the W$_\mathrm{L}$ mode induced here, one has $|k| \le 4~\wce/c$ and thus the right-hand side of the above equation is less than 1.16. According to the energy curves plotted in Figure 1d, the W$_\mathrm{L}$ rises above the noise level around 300 $\wpe^{-1}$, therefore we calculated the value of $T_\mathrm{e\perp}/T_\mathrm{e\parallel}$ at this time and found it is 2.08. This satisfies the above condition, meaning that W$_\mathrm{L}$ is excited by the thermal anisotropy instability due to the ring-distributed electrons.

\subsubsection{The Z mode.}

 The Z mode is the slow extraordinary magnetoionic mode. It is electromagnetic with the magnetic component (mainly $B_z$) much weaker than its electric component (mainly $E_x$), and propagates quasi-perpendicularly with both superluminal and subluminal parts. It reaches the peak level around 500 $\wpe^{-1}$. See Figures 1, 3, {\&} 4. Its range of $\omega$ is (9.7, 10.1) $\wce$ and its range of $|k|$ is (0, 20) $\wce/c$. Two sets of ($\omega, \vec k$) (see arrows in Figure 4 and parameters listed in Table 1) are selected to plot the resonance curves with $n = 10$ in Figure 2b. All curves pass the positive gradient region of VDF, supporting its ECMI origin.

From the above analysis we suggest that the UH1, W$_\mathrm{H}$, and Z modes are excited directly through the kinetic ECMI, and W$_\mathrm{L}$ by the thermal anisotropy instability, while the IA1, IA2, and UH2 modes through the electrostatic decay of the primary UH mode.

\subsubsection{The escaping radiation modes.}

 According to the $\vec k$--space and $\omega$--$k$ spectra (Figures 3 {\&} 4), there exists significant H/F emission. The H emission appears as the circular arcs, propagating quasi-parallel with $\theta \le 45^{\degr}$. It presents a sporadic frequency distribution in the range of (19.2, 20.5) $\wce$. The F emission appears within (10, 10.1) $\wce$ and (0, 2) $\wce/c$. The asymptotic energy of the H emission is about 1.4 $\times 10^{-5}~E_\mathrm{k0}$(A) (see lower panels of Figure 1), much stronger than that of the F emission ($\sim 3\times 10^{-7}~E_\mathrm{k0}$(A)). This (the H emission being much stronger than the F emission) is consistent with the simulation by \citet{2020ApJ...891L..25N} and \citet{2021ApJ...909L...5L} with the DGH distribution of energetic electrons, yet very different from the result obtained for the pure-beam energetic electrons \citep{2022ApJ...924L..34C} in which the O/F emission is comparable to the H emission in intensity. In addition, according to Figure 1 the H emission reaches its asymptotic value at $\sim 400~\wpe^{-1}$, earlier than the F emission ($\sim 600~ \wpe^{-1}$).

The generation of the H emission requires the nonlinear coupling of two modes around $\wpe$. Only the modes stronger than the H emission should be considered. This excludes the Z mode with large--$|k|~(>10~\wce/c)$. The matching conditions of $\vec k$ further exclude the two coalescing processes (UH + Z$\rightarrow$H and Z + Z$\rightarrow$H), the only possibility left is the UH + UH$^\prime$$\rightarrow$H process, where UH$^\prime$ represents the almost-counter propagating UH mode. One can easily verify the satisfaction of the corresponding matching conditions.

To generate the F emission through resonant wave-wave coupling, one needs a high-frequency ($\sim \wpe$) mode and a low-frequency ($< \wce$) mode. The UH and IA1 modes can be rejected immediately according to the matching condition of $k_\perp$, and W$_\mathrm{H}$ mode rejected due to the mismatch of the $\omega$ condition since its frequency is too high. Thus, only the Z mode can act as the high-frequency candidate with IA2 and W$_\mathrm{L}$ as the low-frequency candidates, i.e., in terms of Z+W$_\mathrm{L}$ (or IA2) $\rightarrow$ O/F.

In subsection 3.3, we present Case C with a much larger mass ratio ($m_\mathrm{p}/m_\mathrm{e}=183600$) that is equivalent to assume immobile protons and the IA mode with negligibly-low frequency and intensity. In that case, the O/F emission does not change obviously in frequency ranges in comparison to Case A. This indicates that the IA2 mode does not play a role in the plasma emission process. This leaves only the option of Z + W$_\mathrm{L}$ $\rightarrow$ O/F. Note that the O/F emission is characterized by frequency slightly above its cutoff $\wpe$ and the corresponding small $k$ ($|k| < 2~\wce/c$). In panels b and c of Figure 3, we delineate the $\vec k$ regime of the W$_\mathrm{L}$ mode with dashed circles, which is close to the inner edge of the Z mode regime, indicating the match of the $\vec k$ condition for the two modes coalescing into the small-$|k|$ O/F mode. The $\omega$ matching condition can be satisfied according to the $\omega$ ranges of these modes (see Figure 4).

\subsection{Case B with $u_\mathrm{d\perp} = 0.2c$}

By lowering $u_\perp$ we reduce the overall energy of energetic electrons. This leads to the corresponding variation of the location of positive gradient of the electron VDF (see Figure 2). Two consequences can be expected (1) the modes excited by energetic electrons may become weaker; (2) waves with different values of $(\omega, \vec k)$ will be excited in accordance with the downward shift of the ECMI resonance curves, in other words, efficient excitations of waves will move from those with higher resonance curves to lower ones.

Indeed, the most obvious change of Case B in comparison to Case A is the lowering of mode energy (see Figures 6--7). All relevant modes including the non-escaping (UH, Z, W, and IA) and escaping (the F/H plasma emissions) ones exhibit weaker intensity. This is also observed from the energy curves of various field components and wave modes (Figure 1). Note that the relative peak magnitude of the $E_x$ energy does not show considerable change from Case A to Case B, while the $B_y$ peak energy decreases from $\sim 2.5 \times 10^{-3}~E_\mathrm{k0}$(A) to $\sim 7 \times 10^{-4}~E_\mathrm{k0}$(B) where $ E_\mathrm{k0}\mathrm{(B)}~(= 1.655~m_\mathrm{e0}c^2)$ is the total initial kinetic energy of the ring electrons in Case B; and the energy maxima of the UH and W$_\mathrm{H}$ modes decrease from $\sim 6 \times 10^{-2}$ and $\sim 2.5 \times 10^{-3}~E_\mathrm{k0}$(A) to $\sim 5 \times 10^{-2}$ and $\sim 1 \times 10^{-6} ~ E_\mathrm{k0}$(B), respectively. Such declines are mainly due to the decrease of free energy carried by the energetic electrons.

In addition to the energy changes, the $k_\perp$ of the UH mode (including both UH1 and UH2) increases from $\sim$ 50 to $\sim$ 65 $\wce/c$. The IA1 mode also increases in $k_\perp$ from 100 to 130 $\wce/c$, keeping to be two times of $k_\perp$ of the UH mode, while the IA2 mode ($k_\perp \approx 0 $) is hardly observable (see the movie accompanying Figure 5). In addition, the $k_\parallel$ range of IA1 decreases from (0, $\pm 50$) $\wce/c$ to less than (0, $\pm 20$) $\wce/c$, consistent with the corresponding shrinkage of the $k_\parallel$ range of the UH mode. These observations are in line with the UH1--decay scenario for the generation of UH2, IA1, and IA2. In Case B only the quasi-parallel W$_\mathrm{H}$ mode ($\theta \approx 0$) with a much weaker intensity is excited, while in Case A it can be excited over a wider range of $\theta$. The Z mode does not present efficient excitation below $\omega < 9.8~\wce$, while in Case A such excitation exists.

In Figure 2 (lower panels), we present resonance curves for the UH1, W$_\mathrm{H}$, and Z modes (see arrows in Figure 7 for their spectral locations and parameters listed in Table 1). The resonance curve of the Z mode with $\omega = 9.72$ $\wce$ does not pass through the positive-gradient region of the VDF, thus no significant excitation is observed from Figure 7b. For the UH1 mode, again the resonance curve of UH1 passes through the region with significant positive gradient. This agrees with the observation from Figures 6 and 7 that UH1 is the dominant mode. The curve of UH2 passes through the strong negative-gradient region of the background VDF, meaning that it cannot be excited via the ECMI and its damping is due to thermal absorption. For the W$_\mathrm{H}$ mode, the resonance curve only passes through the edge of the electron VDF (see Figure 2h) thus its corresponding intensity is much weaker than that in Case A. For the W$_\mathrm{L}$ mode, again the resonance curve is too large in coordinates to be visible; the corresponding value of $T_\perp / T_\parallel$ is 1.59, also satisfying the condition of the thermal anisotropic instability (Equation 3). This reveals its origin.

By analyzing the ($\omega, \vec k$) ranges of relevant modes, we infer that the matching conditions of the proposed plasma emission process (UH + UH$^\prime~\rightarrow$ H and Z + W$_\mathrm{L}~\rightarrow$ O/F) can be satisfied. In Figure 6c, we delineated the $\omega-\vec k$ regime of the W$_\mathrm{L}$ mode with a dashed circle, which has been overplotted in the Z mode dispersion diagram (see Figure 6b) for comparison. The wavenumber ranges of the two modes are close to each other with overlaps. In addition, according to Figure 7 the frequency of the W$_\mathrm{L}$ modes is $\le$ 0.1 $\wce$, while the frequency of the Z mode is from 9.8 $\wce$ to slightly above 10 $\Omega_{ce}$. Therefore, the two modes can coalesce to give the O/F mode at frequency around 10 $\wce$ with a wavenumber less than 2 $\wce/c$.

Comparing the mode energy plots (Figure 1) for Case B with Case A, the Z-mode energy presents a slight increase, and W$_\mathrm{H}$ presents a large drop while both W$_\mathrm{L}$ and O/F decline slightly in relative energy. These observations are not against the proposed radiation process. The UH mode does not change considerably in relative intensity normalized by the corresponding $E_\mathrm{k0}$, while the H emission declines significantly. This is due to the increase of $k_\perp$ of the UH mode that greatly limits ranges of the UH mode participating the three-wave resonance.

\subsection{Case C with $u_\mathrm{d\perp} = 0.3c$ and $m_\mathrm{p}/m_\mathrm{e}=183600$}

This case is a numerical experiment to see how various modes change according to $m_\mathrm{p}/m_\mathrm{e}$ to reveal more clues on wave excitation. See Figures 8 and 9 for the $\vec k$-space wave map and the $\omega-k$ dispersion diagram. In comparison with Case A, the UH mode does not change much in both energy and intensity, neither does the H mode; in addition, the frequency (and intensity) of IA becomes negligible while the Z, W, and O/F modes do not change much in both intensities and frequency ranges. This gives the earlier deduction that IA2 does not contribute to the F emission. In addition, both Z and O/F modes present weak and comparable enhancement according to their energy profiles plotted in Figure 1 (see also Figure 9). The energies of all other modes including the UH mode and the H emission do not change considerably from Case A to Case C. These observations are not inconsistent with the proposed plasma radiation mechanism.

\section{Conclusions and discussion}

Using fully-kinetic electromagnetic particle-in-cell simulations, we investigate the physics of wave excitation and plasma emission driven by the ring-distributed energetic electrons in overdense plasmas of the solar coronal conditions. The primary modes include the upper hybrid (UH) mode and the whistler mode; other modes include the Z and ion-acoustic (IA) mode as well as the escaping fundamental (F) and harmonic (H) plasma emission. We infer that (1) the primary UH mode, the high-frequency W mode (W$_\mathrm{H}$), and the Z mode are excited through the electron cyclotron maser instability (ECMI), and the secondary (quasi-perpendicular) UH mode and the IA mode generated by the decay process of the primary UH mode, while the low-frequency W mode (W$_\mathrm{L}$) is driven by the thermal-anisotropy instability of the ring-plasma system; (2) the F emission is generated by the nonlinear coupling of the Z mode with W$_\mathrm{L}$ (i.e., Z + W$_\mathrm{L}~\rightarrow$ O/F), and the H emission is generated by the nonlinear coupling of the almost counter-propagating UH modes (i.e., UH + UH$^\prime~\rightarrow$ H).

The above radiation process is different from that described by the standard scenario of plasma emission which starts from the kinetic beam-driven bump-on-tail instability and further nonlinear couplings of the enhanced Langmuir wave and other secondary modes such as the IA mode for the F emission and the scattered backward-propagating Langmuir mode for the H emission. The standard scenario has been verified lately by many authors using fully-kinetic electromagnetic PIC simulations, such as \citet{2015A&A...584A..83T}, \citet{2019JGRA..124.1475H} and \citet{2022arXiv220911707Z} for unmagnetized plasmas, and \citet{2022ApJ...924L..34C} for weakly-magnetized or overdense plasmas. \citet{2020ApJ...891L..25N} simulated the radiation process within overdense plasmas interacting with energetic electrons of a double-sided loss-cone distribution (i.e., the DGH distribution) and concluded that the F emission is generated by the almost counter-propagating Z and W modes and the H emission by the almost counter-propagating UH modes, consistent with the findings reported here for the ring-plasma system. Combining the present study with that of \citet{2020ApJ...891L..25N}, we suggest that there exist two parallel mechanisms of plasma emission in overdense plasmas, one is driven by the beam-like energetic electrons with free energy dominated by their parallel motion (i.e., $\partial f / \partial v_\parallel > 0$) according to the standard scenario, the other is driven by the ring/DGH/loss-cone/horseshoe-like energetic electrons with free energy dominated by their perpendicular motion (i.e., $\partial f / \partial v_\perp > 0$).

One major difference between the pure-beam simulation \citep{2022ApJ...924L..34C} and the pure-ring simulation presented here is the relative intensity of the H emission and the O/F emission. For the pure-beam case the two emissions are comparable in energy, while for the pure-ring case the H emission is much stronger than the F emission (by 1-2 orders in magnitude). Simulations of energetic electrons with the DGH distribution \citep{2020ApJ...891L..25N,2021ApJ...909L...5L} lead to results similar to the pure-ring case with the H emission being much stronger than the F emission. Other differences include, (1) the primary mode in the pure-beam case is the parallel-to-obliquely propagating beam-Langmuir wave excited by the kinetic bump-on-tail instability, while it is the UH mode (i.e., the obliquely-to-perpendicular propagating Langmuir wave) driven by the ECMI in the pure-ring (and DGH) cases; (2) the frequency distribution, relative intensity, and excitation mechanism of the whistler mode are also different for different cases, for the pure-ring case there exist two different excitation mechanisms including the ECMI and the anisotropic thermal instability, resulting in the two-component whistler wave, while it is mainly excited by the ECMI in the pure-beam case; and (3) the excitation mechanism of the electromagnetic Z mode wave is attributed to the decay of the beam-Langmuir wave in the pure-beam case and to direct ECMI excitation in the pure-ring case. These conclusions should be taken into account for future observational and theoretical studies on coherent plasma radiation in space and astrophysical plasmas.

\begin{acknowledgments}
This study is supported by NNSFC grants (11790303 (11790300), 11973031, and 11873036). The authors acknowledge Dr. Quanming Lu, Xinliang Gao, and Xiaocan Li for helpful discussion, the Beijing Super Cloud Computing Center (BSC-C, URL: http://www.blsc.cn/) for computational resources, and LANL for the open-source VPIC code.
\end{acknowledgments}

\section*{Data Availability Statement}

The data that support the findings of this study are available from the corresponding author upon reasonable request.

\nocite{*}
\bibliographystyle{aipnum4-1.bst}
\bibliography{Chen2022POP}

\begin{thebibliography}{68}%
\makeatletter
\providecommand \@ifxundefined [1]{%
 \@ifx{#1\undefined}
}%
\providecommand \@ifnum [1]{%
 \ifnum #1\expandafter \@firstoftwo
 \else \expandafter \@secondoftwo
 \fi
}%
\providecommand \@ifx [1]{%
 \ifx #1\expandafter \@firstoftwo
 \else \expandafter \@secondoftwo
 \fi
}%
\providecommand \natexlab [1]{#1}%
\providecommand \enquote  [1]{``#1''}%
\providecommand \bibnamefont  [1]{#1}%
\providecommand \bibfnamefont [1]{#1}%
\providecommand \citenamefont [1]{#1}%
\providecommand \href@noop [0]{\@secondoftwo}%
\providecommand \href [0]{\begingroup \@sanitize@url \@href}%
\providecommand \@href[1]{\@@startlink{#1}\@@href}%
\providecommand \@@href[1]{\endgroup#1\@@endlink}%
\providecommand \@sanitize@url [0]{\catcode `\\12\catcode `\$12\catcode
  `\&12\catcode `\#12\catcode `\^12\catcode `\_12\catcode `\%12\relax}%
\providecommand \@@startlink[1]{}%
\providecommand \@@endlink[0]{}%
\providecommand \url  [0]{\begingroup\@sanitize@url \@url }%
\providecommand \@url [1]{\endgroup\@href {#1}{\urlprefix }}%
\providecommand \urlprefix  [0]{URL }%
\providecommand \Eprint [0]{\href }%
\providecommand \doibase [0]{http://dx.doi.org/}%
\providecommand \selectlanguage [0]{\@gobble}%
\providecommand \bibinfo  [0]{\@secondoftwo}%
\providecommand \bibfield  [0]{\@secondoftwo}%
\providecommand \translation [1]{[#1]}%
\providecommand \BibitemOpen [0]{}%
\providecommand \bibitemStop [0]{}%
\providecommand \bibitemNoStop [0]{.\EOS\space}%
\providecommand \EOS [0]{\spacefactor3000\relax}%
\providecommand \BibitemShut  [1]{\csname bibitem#1\endcsname}%
\let\auto@bib@innerbib\@empty
\bibitem [{\citenamefont {{Ginzburg}}\ and\ \citenamefont
  {{Zhelezniakov}}(1958)}]{1958SvA.....2..653G}%
  \BibitemOpen
  \bibfield  {author} {\bibinfo {author} {\bibfnamefont {V.~L.}\ \bibnamefont
  {{Ginzburg}}}\ and\ \bibinfo {author} {\bibfnamefont {V.~V.}\ \bibnamefont
  {{Zhelezniakov}}},\ }\href@noop {} {\bibfield  {journal} {\bibinfo  {journal}
  {Sov. Astron.}\ }\textbf {\bibinfo {volume} {2}},\ \bibinfo {pages} {653}
  (\bibinfo {year} {1958})}\BibitemShut {NoStop}%
\bibitem [{\citenamefont {{Wild}}(1950{\natexlab{a}})}]{1950AuSRA...3..399W}%
  \BibitemOpen
  \bibfield  {author} {\bibinfo {author} {\bibfnamefont {J.~P.}\ \bibnamefont
  {{Wild}}},\ }\href {\doibase 10.1071/CH9500399} {\bibfield  {journal}
  {\bibinfo  {journal} {Australian Journal of Scientific Research A Physical
  Sciences}\ }\textbf {\bibinfo {volume} {3}},\ \bibinfo {pages} {399}
  (\bibinfo {year} {1950}{\natexlab{a}})}\BibitemShut {NoStop}%
\bibitem [{\citenamefont {{Wild}}(1950{\natexlab{b}})}]{1950AuSRA...3..541W}%
  \BibitemOpen
  \bibfield  {author} {\bibinfo {author} {\bibfnamefont {J.~P.}\ \bibnamefont
  {{Wild}}},\ }\href {\doibase 10.1071/CH9500541} {\bibfield  {journal}
  {\bibinfo  {journal} {Australian Journal of Scientific Research A Physical
  Sciences}\ }\textbf {\bibinfo {volume} {3}},\ \bibinfo {pages} {541}
  (\bibinfo {year} {1950}{\natexlab{b}})}\BibitemShut {NoStop}%
\bibitem [{\citenamefont {{Wild}}, \citenamefont {{Murray}},\ and\
  \citenamefont {{Rowe}}(1954)}]{1954AuJPh...7..439W}%
  \BibitemOpen
  \bibfield  {author} {\bibinfo {author} {\bibfnamefont {J.~P.}\ \bibnamefont
  {{Wild}}}, \bibinfo {author} {\bibfnamefont {J.~D.}\ \bibnamefont
  {{Murray}}}, \ and\ \bibinfo {author} {\bibfnamefont {W.~C.}\ \bibnamefont
  {{Rowe}}},\ }\href {\doibase 10.1071/PH540439} {\bibfield  {journal}
  {\bibinfo  {journal} {Aust. J. Phys.}\ }\textbf {\bibinfo {volume} {7}},\
  \bibinfo {pages} {439} (\bibinfo {year} {1954})}\BibitemShut {NoStop}%
\bibitem [{\citenamefont {{McLean}}\ and\ \citenamefont
  {{Labrum}}(1985)}]{1985srph.book.....M}%
  \BibitemOpen
  \bibfield  {author} {\bibinfo {author} {\bibfnamefont {D.~J.}\ \bibnamefont
  {{McLean}}}\ and\ \bibinfo {author} {\bibfnamefont {N.~R.}\ \bibnamefont
  {{Labrum}}},\ }\href@noop {} {\emph {\bibinfo {title} {{Solar radiophysics :
  studies of emission from the sun at metre wavelengths}}}}\ (\bibinfo
  {publisher} {Cambridge; New York : Cambridge University Press},\ \bibinfo
  {year} {1985})\BibitemShut {NoStop}%
\bibitem [{\citenamefont {{Melrose}}(1970)}]{1970AuJPh..23..871M}%
  \BibitemOpen
  \bibfield  {author} {\bibinfo {author} {\bibfnamefont {D.~B.}\ \bibnamefont
  {{Melrose}}},\ }\href {\doibase 10.1071/PH700871} {\bibfield  {journal}
  {\bibinfo  {journal} {Aust. J. Phys.}\ }\textbf {\bibinfo {volume} {23}},\
  \bibinfo {pages} {871} (\bibinfo {year} {1970})}\BibitemShut {NoStop}%
\bibitem [{\citenamefont {{Takakura}}(1979)}]{1979SoPh...61..161T}%
  \BibitemOpen
  \bibfield  {author} {\bibinfo {author} {\bibfnamefont {T.}~\bibnamefont
  {{Takakura}}},\ }\href {\doibase 10.1007/BF00155454} {\bibfield  {journal}
  {\bibinfo  {journal} {Sol. Phys.}\ }\textbf {\bibinfo {volume} {61}},\
  \bibinfo {pages} {161} (\bibinfo {year} {1979})}\BibitemShut {NoStop}%
\bibitem [{\citenamefont {{Cairns}}\ and\ \citenamefont
  {{Melrose}}(1985)}]{1985JGR....90.6637C}%
  \BibitemOpen
  \bibfield  {author} {\bibinfo {author} {\bibfnamefont {I.~H.}\ \bibnamefont
  {{Cairns}}}\ and\ \bibinfo {author} {\bibfnamefont {D.~B.}\ \bibnamefont
  {{Melrose}}},\ }\href {\doibase 10.1029/JA090iA07p06637} {\bibfield
  {journal} {\bibinfo  {journal} {J. Geophys. Res.}\ }\textbf {\bibinfo
  {volume} {90}},\ \bibinfo {pages} {6637} (\bibinfo {year}
  {1985})}\BibitemShut {NoStop}%
\bibitem [{\citenamefont {{Cairns}}(1987)}]{1987JPlPh..38..169C}%
  \BibitemOpen
  \bibfield  {author} {\bibinfo {author} {\bibfnamefont {I.~H.}\ \bibnamefont
  {{Cairns}}},\ }\href {\doibase 10.1017/S0022377800012496} {\bibfield
  {journal} {\bibinfo  {journal} {Journal of Plasma Physics}\ }\textbf
  {\bibinfo {volume} {38}},\ \bibinfo {pages} {169} (\bibinfo {year}
  {1987})}\BibitemShut {NoStop}%
\bibitem [{\citenamefont {{Cairns}}(1988)}]{1988JGR....93.3958C}%
  \BibitemOpen
  \bibfield  {author} {\bibinfo {author} {\bibfnamefont {I.~H.}\ \bibnamefont
  {{Cairns}}},\ }\href {\doibase 10.1029/JA093iA05p03958} {\bibfield  {journal}
  {\bibinfo  {journal} {J. Geophys. Res.}\ }\textbf {\bibinfo {volume} {93}},\
  \bibinfo {pages} {3958} (\bibinfo {year} {1988})}\BibitemShut {NoStop}%
\bibitem [{\citenamefont {{Robinson}}, \citenamefont {{Willes}},\ and\
  \citenamefont {{Cairns}}(1993)}]{1993ApJ...408..720R}%
  \BibitemOpen
  \bibfield  {author} {\bibinfo {author} {\bibfnamefont {P.~A.}\ \bibnamefont
  {{Robinson}}}, \bibinfo {author} {\bibfnamefont {A.~J.}\ \bibnamefont
  {{Willes}}}, \ and\ \bibinfo {author} {\bibfnamefont {I.~H.}\ \bibnamefont
  {{Cairns}}},\ }\href {\doibase 10.1086/172632} {\bibfield  {journal}
  {\bibinfo  {journal} {\apj}\ }\textbf {\bibinfo {volume} {408}},\ \bibinfo
  {pages} {720} (\bibinfo {year} {1993})}\BibitemShut {NoStop}%
\bibitem [{\citenamefont {{Robinson}}, \citenamefont {{Cairns}},\ and\
  \citenamefont {{Willes}}(1994)}]{1994ApJ...422..870R}%
  \BibitemOpen
  \bibfield  {author} {\bibinfo {author} {\bibfnamefont {P.~A.}\ \bibnamefont
  {{Robinson}}}, \bibinfo {author} {\bibfnamefont {I.~H.}\ \bibnamefont
  {{Cairns}}}, \ and\ \bibinfo {author} {\bibfnamefont {A.~J.}\ \bibnamefont
  {{Willes}}},\ }\href {\doibase 10.1086/173779} {\bibfield  {journal}
  {\bibinfo  {journal} {\apj}\ }\textbf {\bibinfo {volume} {422}},\ \bibinfo
  {pages} {870} (\bibinfo {year} {1994})}\BibitemShut {NoStop}%
\bibitem [{\citenamefont {{Melrose}}(1980)}]{1980SSRv...26....3M}%
  \BibitemOpen
  \bibfield  {author} {\bibinfo {author} {\bibfnamefont {D.~B.}\ \bibnamefont
  {{Melrose}}},\ }\href {\doibase 10.1007/BF00212597} {\bibfield  {journal}
  {\bibinfo  {journal} {Space Science Reviews}\ }\textbf {\bibinfo {volume}
  {26}},\ \bibinfo {pages} {3} (\bibinfo {year} {1980})}\BibitemShut {NoStop}%
\bibitem [{\citenamefont {{Melrose}}(1986)}]{1986islp.book.....M}%
  \BibitemOpen
  \bibfield  {author} {\bibinfo {author} {\bibfnamefont {D.~B.}\ \bibnamefont
  {{Melrose}}},\ }\href@noop {} {\emph {\bibinfo {title} {{Instabilities in
  Space and Laboratory Plasmas}}}}\ (\bibinfo  {publisher} {Cambridge, UK:
  Cambridge University Press},\ \bibinfo {year} {1986})\BibitemShut {NoStop}%
\bibitem [{\citenamefont {{Melrose}}(2017)}]{2017RvMPP...1....5M}%
  \BibitemOpen
  \bibfield  {author} {\bibinfo {author} {\bibfnamefont {D.~B.}\ \bibnamefont
  {{Melrose}}},\ }\href {\doibase 10.1007/s41614-017-0007-0} {\bibfield
  {journal} {\bibinfo  {journal} {Reviews of Modern Plasma Physics}\ }\textbf
  {\bibinfo {volume} {1}},\ \bibinfo {eid} {5} (\bibinfo {year} {2017})},\
  \Eprint {http://arxiv.org/abs/1707.02009} {arXiv:1707.02009
  [physics.plasm-ph]} \BibitemShut {NoStop}%
\bibitem [{\citenamefont {{Hinkel-Lipsker}}, \citenamefont {{Fried}},\ and\
  \citenamefont {{Morales}}(1992)}]{1992PhFlB...4.1772H}%
  \BibitemOpen
  \bibfield  {author} {\bibinfo {author} {\bibfnamefont {D.~E.}\ \bibnamefont
  {{Hinkel-Lipsker}}}, \bibinfo {author} {\bibfnamefont {B.~D.}\ \bibnamefont
  {{Fried}}}, \ and\ \bibinfo {author} {\bibfnamefont {G.~J.}\ \bibnamefont
  {{Morales}}},\ }\href {\doibase 10.1063/1.860033} {\bibfield  {journal}
  {\bibinfo  {journal} {Physics of Fluids B}\ }\textbf {\bibinfo {volume}
  {4}},\ \bibinfo {pages} {1772} (\bibinfo {year} {1992})}\BibitemShut
  {NoStop}%
\bibitem [{\citenamefont {{Kim}}, \citenamefont {{Cairns}},\ and\ \citenamefont
  {{Robinson}}(2007)}]{2007PhRvL..99a5003K}%
  \BibitemOpen
  \bibfield  {author} {\bibinfo {author} {\bibfnamefont {E.-H.}\ \bibnamefont
  {{Kim}}}, \bibinfo {author} {\bibfnamefont {I.~H.}\ \bibnamefont {{Cairns}}},
  \ and\ \bibinfo {author} {\bibfnamefont {P.~A.}\ \bibnamefont {{Robinson}}},\
  }\href {\doibase 10.1103/PhysRevLett.99.015003} {\bibfield  {journal}
  {\bibinfo  {journal} {\prl}\ }\textbf {\bibinfo {volume} {99}},\ \bibinfo
  {eid} {015003} (\bibinfo {year} {2007})}\BibitemShut {NoStop}%
\bibitem [{\citenamefont {{Yoon}}\ \emph {et~al.}(1994)\citenamefont {{Yoon}},
  \citenamefont {{Wu}}, \citenamefont {{Vinas}}, \citenamefont {{Reiner}},
  \citenamefont {{Fainberg}},\ and\ \citenamefont
  {{Stone}}}]{1994JGR....9923481Y}%
  \BibitemOpen
  \bibfield  {author} {\bibinfo {author} {\bibfnamefont {P.~H.}\ \bibnamefont
  {{Yoon}}}, \bibinfo {author} {\bibfnamefont {C.~S.}\ \bibnamefont {{Wu}}},
  \bibinfo {author} {\bibfnamefont {A.~F.}\ \bibnamefont {{Vinas}}}, \bibinfo
  {author} {\bibfnamefont {M.~J.}\ \bibnamefont {{Reiner}}}, \bibinfo {author}
  {\bibfnamefont {J.}~\bibnamefont {{Fainberg}}}, \ and\ \bibinfo {author}
  {\bibfnamefont {R.~G.}\ \bibnamefont {{Stone}}},\ }\href {\doibase
  10.1029/94JA02489} {\bibfield  {journal} {\bibinfo  {journal} {J. Geophys.
  Res.}\ }\textbf {\bibinfo {volume} {99}},\ \bibinfo {pages} {23,481}
  (\bibinfo {year} {1994})}\BibitemShut {NoStop}%
\bibitem [{\citenamefont {{Papadopoulos}}\ and\ \citenamefont
  {{Freund}}(1978)}]{1978GeoRL...5..881P}%
  \BibitemOpen
  \bibfield  {author} {\bibinfo {author} {\bibfnamefont {K.}~\bibnamefont
  {{Papadopoulos}}}\ and\ \bibinfo {author} {\bibfnamefont {H.~P.}\
  \bibnamefont {{Freund}}},\ }\href {\doibase 10.1029/GL005i010p00881}
  {\bibfield  {journal} {\bibinfo  {journal} {Geophys. Res. Lett.}\ }\textbf
  {\bibinfo {volume} {5}},\ \bibinfo {pages} {881} (\bibinfo {year}
  {1978})}\BibitemShut {NoStop}%
\bibitem [{\citenamefont {{Goldman}}, \citenamefont {{Reiter}},\ and\
  \citenamefont {{Nicholson}}(1980)}]{1980PhFl...23..388G}%
  \BibitemOpen
  \bibfield  {author} {\bibinfo {author} {\bibfnamefont {M.~V.}\ \bibnamefont
  {{Goldman}}}, \bibinfo {author} {\bibfnamefont {G.~F.}\ \bibnamefont
  {{Reiter}}}, \ and\ \bibinfo {author} {\bibfnamefont {D.~R.}\ \bibnamefont
  {{Nicholson}}},\ }\href {\doibase 10.1063/1.862982} {\bibfield  {journal}
  {\bibinfo  {journal} {Physics of Fluids}\ }\textbf {\bibinfo {volume} {23}},\
  \bibinfo {pages} {388} (\bibinfo {year} {1980})}\BibitemShut {NoStop}%
\bibitem [{\citenamefont {{Malaspina}}, \citenamefont {{Cairns}},\ and\
  \citenamefont {{Ergun}}(2010)}]{2010JGRA..115.1101M}%
  \BibitemOpen
  \bibfield  {author} {\bibinfo {author} {\bibfnamefont {D.~M.}\ \bibnamefont
  {{Malaspina}}}, \bibinfo {author} {\bibfnamefont {I.~H.}\ \bibnamefont
  {{Cairns}}}, \ and\ \bibinfo {author} {\bibfnamefont {R.~E.}\ \bibnamefont
  {{Ergun}}},\ }\href {\doibase 10.1029/2009JA014609} {\bibfield  {journal}
  {\bibinfo  {journal} {Journal of Geophysical Research (Space Physics)}\
  }\textbf {\bibinfo {volume} {115}},\ \bibinfo {eid} {A01101} (\bibinfo {year}
  {2010})}\BibitemShut {NoStop}%
\bibitem [{\citenamefont {{Malaspina}}, \citenamefont {{Cairns}},\ and\
  \citenamefont {{Ergun}}(2011)}]{2011GeoRL..3813101M}%
  \BibitemOpen
  \bibfield  {author} {\bibinfo {author} {\bibfnamefont {D.~M.}\ \bibnamefont
  {{Malaspina}}}, \bibinfo {author} {\bibfnamefont {I.~H.}\ \bibnamefont
  {{Cairns}}}, \ and\ \bibinfo {author} {\bibfnamefont {R.~E.}\ \bibnamefont
  {{Ergun}}},\ }\href {\doibase 10.1029/2011GL047642} {\bibfield  {journal}
  {\bibinfo  {journal} {Geophys. Res. Lett.}\ }\textbf {\bibinfo {volume}
  {38}},\ \bibinfo {eid} {L13101} (\bibinfo {year} {2011})}\BibitemShut
  {NoStop}%
\bibitem [{\citenamefont {{Malaspina}}, \citenamefont {{Cairns}},\ and\
  \citenamefont {{Ergun}}(2012)}]{2012ApJ...755...45M}%
  \BibitemOpen
  \bibfield  {author} {\bibinfo {author} {\bibfnamefont {D.~M.}\ \bibnamefont
  {{Malaspina}}}, \bibinfo {author} {\bibfnamefont {I.~H.}\ \bibnamefont
  {{Cairns}}}, \ and\ \bibinfo {author} {\bibfnamefont {R.~E.}\ \bibnamefont
  {{Ergun}}},\ }\href {\doibase 10.1088/0004-637X/755/1/45} {\bibfield
  {journal} {\bibinfo  {journal} {\apj}\ }\textbf {\bibinfo {volume} {755}},\
  \bibinfo {eid} {45} (\bibinfo {year} {2012})}\BibitemShut {NoStop}%
\bibitem [{\citenamefont {{Malaspina}}\ \emph {et~al.}(2013)\citenamefont
  {{Malaspina}}, \citenamefont {{Graham}}, \citenamefont {{Ergun}},\ and\
  \citenamefont {{Cairns}}}]{2013JGRA..118.6880M}%
  \BibitemOpen
  \bibfield  {author} {\bibinfo {author} {\bibfnamefont {D.~M.}\ \bibnamefont
  {{Malaspina}}}, \bibinfo {author} {\bibfnamefont {D.~B.}\ \bibnamefont
  {{Graham}}}, \bibinfo {author} {\bibfnamefont {R.~E.}\ \bibnamefont
  {{Ergun}}}, \ and\ \bibinfo {author} {\bibfnamefont {I.~H.}\ \bibnamefont
  {{Cairns}}},\ }\href {\doibase 10.1002/2013JA019309} {\bibfield  {journal}
  {\bibinfo  {journal} {Journal of Geophysical Research (Space Physics)}\
  }\textbf {\bibinfo {volume} {118}},\ \bibinfo {pages} {6880} (\bibinfo {year}
  {2013})}\BibitemShut {NoStop}%
\bibitem [{\citenamefont {{Chiu}}(1970)}]{1970SoPh...13..420C}%
  \BibitemOpen
  \bibfield  {author} {\bibinfo {author} {\bibfnamefont {Y.~T.}\ \bibnamefont
  {{Chiu}}},\ }\href {\doibase 10.1007/BF00153561} {\bibfield  {journal}
  {\bibinfo  {journal} {Sol. Phys.}\ }\textbf {\bibinfo {volume} {13}},\
  \bibinfo {pages} {420} (\bibinfo {year} {1970})}\BibitemShut {NoStop}%
\bibitem [{\citenamefont {{Chin}}(1972)}]{1972P&SS...20..711C}%
  \BibitemOpen
  \bibfield  {author} {\bibinfo {author} {\bibfnamefont {Y.-C.}\ \bibnamefont
  {{Chin}}},\ }\href {\doibase 10.1016/0032-0633(72)90155-9} {\bibfield
  {journal} {\bibinfo  {journal} {Planet. Space Sci.}\ }\textbf {\bibinfo
  {volume} {20}},\ \bibinfo {pages} {711} (\bibinfo {year} {1972})}\BibitemShut
  {NoStop}%
\bibitem [{\citenamefont {{Melrose}}(1975)}]{1975AuJPh..28..101M}%
  \BibitemOpen
  \bibfield  {author} {\bibinfo {author} {\bibfnamefont {D.~B.}\ \bibnamefont
  {{Melrose}}},\ }\href {\doibase 10.1071/PH750101} {\bibfield  {journal}
  {\bibinfo  {journal} {Aust. J. Phys.}\ }\textbf {\bibinfo {volume} {28}},\
  \bibinfo {pages} {101} (\bibinfo {year} {1975})}\BibitemShut {NoStop}%
\bibitem [{\citenamefont {{Ni}}\ \emph {et~al.}(2021)\citenamefont {{Ni}},
  \citenamefont {{Chen}}, \citenamefont {{Li}}, \citenamefont {{Sun}},
  \citenamefont {{Ning}},\ and\ \citenamefont {{Zhang}}}]{2021PhPl...28d0701N}%
  \BibitemOpen
  \bibfield  {author} {\bibinfo {author} {\bibfnamefont {S.}~\bibnamefont
  {{Ni}}}, \bibinfo {author} {\bibfnamefont {Y.}~\bibnamefont {{Chen}}},
  \bibinfo {author} {\bibfnamefont {C.}~\bibnamefont {{Li}}}, \bibinfo {author}
  {\bibfnamefont {J.}~\bibnamefont {{Sun}}}, \bibinfo {author} {\bibfnamefont
  {H.}~\bibnamefont {{Ning}}}, \ and\ \bibinfo {author} {\bibfnamefont
  {Z.}~\bibnamefont {{Zhang}}},\ }\href {\doibase 10.1063/5.0045546} {\bibfield
   {journal} {\bibinfo  {journal} {Physics of Plasmas}\ }\textbf {\bibinfo
  {volume} {28}},\ \bibinfo {eid} {040701} (\bibinfo {year} {2021})},\ \Eprint
  {http://arxiv.org/abs/2104.04267} {arXiv:2104.04267 [physics.plasm-ph]}
  \BibitemShut {NoStop}%
\bibitem [{\citenamefont {{Kasaba}}, \citenamefont {{Matsumoto}},\ and\
  \citenamefont {{Omura}}(2001)}]{2001JGR...10618693K}%
  \BibitemOpen
  \bibfield  {author} {\bibinfo {author} {\bibfnamefont {Y.}~\bibnamefont
  {{Kasaba}}}, \bibinfo {author} {\bibfnamefont {H.}~\bibnamefont
  {{Matsumoto}}}, \ and\ \bibinfo {author} {\bibfnamefont {Y.}~\bibnamefont
  {{Omura}}},\ }\href {\doibase 10.1029/2000JA000329} {\bibfield  {journal}
  {\bibinfo  {journal} {J. Geophys. Res.}\ }\textbf {\bibinfo {volume} {106}},\
  \bibinfo {pages} {18693} (\bibinfo {year} {2001})}\BibitemShut {NoStop}%
\bibitem [{\citenamefont {{Rhee}}\ \emph {et~al.}(2009)\citenamefont {{Rhee}},
  \citenamefont {{Ryu}}, \citenamefont {{Woo}}, \citenamefont {{Kaang}},
  \citenamefont {{Yi}},\ and\ \citenamefont {{Yoon}}}]{2009ApJ...694..618R}%
  \BibitemOpen
  \bibfield  {author} {\bibinfo {author} {\bibfnamefont {T.}~\bibnamefont
  {{Rhee}}}, \bibinfo {author} {\bibfnamefont {C.-M.}\ \bibnamefont {{Ryu}}},
  \bibinfo {author} {\bibfnamefont {M.}~\bibnamefont {{Woo}}}, \bibinfo
  {author} {\bibfnamefont {H.~H.}\ \bibnamefont {{Kaang}}}, \bibinfo {author}
  {\bibfnamefont {S.}~\bibnamefont {{Yi}}}, \ and\ \bibinfo {author}
  {\bibfnamefont {P.~H.}\ \bibnamefont {{Yoon}}},\ }\href {\doibase
  10.1088/0004-637X/694/1/618} {\bibfield  {journal} {\bibinfo  {journal}
  {\apj}\ }\textbf {\bibinfo {volume} {694}},\ \bibinfo {pages} {618} (\bibinfo
  {year} {2009})}\BibitemShut {NoStop}%
\bibitem [{\citenamefont {{Umeda}}(2010)}]{2010JGRA..115.1204U}%
  \BibitemOpen
  \bibfield  {author} {\bibinfo {author} {\bibfnamefont {T.}~\bibnamefont
  {{Umeda}}},\ }\href {\doibase 10.1029/2009JA014643} {\bibfield  {journal}
  {\bibinfo  {journal} {Journal of Geophysical Research (Space Physics)}\
  }\textbf {\bibinfo {volume} {115}},\ \bibinfo {eid} {A01204} (\bibinfo {year}
  {2010})}\BibitemShut {NoStop}%
\bibitem [{\citenamefont {{Tsiklauri}}(2011)}]{2011PhPl...18e2903T}%
  \BibitemOpen
  \bibfield  {author} {\bibinfo {author} {\bibfnamefont {D.}~\bibnamefont
  {{Tsiklauri}}},\ }\href {\doibase 10.1063/1.3590928} {\bibfield  {journal}
  {\bibinfo  {journal} {Physics of Plasmas}\ }\textbf {\bibinfo {volume}
  {18}},\ \bibinfo {pages} {052903} (\bibinfo {year} {2011})},\ \Eprint
  {http://arxiv.org/abs/1011.5832} {arXiv:1011.5832 [astro-ph.SR]} \BibitemShut
  {NoStop}%
\bibitem [{\citenamefont {{Ganse}}\ \emph {et~al.}(2012)\citenamefont
  {{Ganse}}, \citenamefont {{Kilian}}, \citenamefont {{Vainio}},\ and\
  \citenamefont {{Spanier}}}]{2012SoPh..280..551G}%
  \BibitemOpen
  \bibfield  {author} {\bibinfo {author} {\bibfnamefont {U.}~\bibnamefont
  {{Ganse}}}, \bibinfo {author} {\bibfnamefont {P.}~\bibnamefont {{Kilian}}},
  \bibinfo {author} {\bibfnamefont {R.}~\bibnamefont {{Vainio}}}, \ and\
  \bibinfo {author} {\bibfnamefont {F.}~\bibnamefont {{Spanier}}},\ }\href
  {\doibase 10.1007/s11207-012-0077-7} {\bibfield  {journal} {\bibinfo
  {journal} {Sol. Phys.}\ }\textbf {\bibinfo {volume} {280}},\ \bibinfo {pages}
  {551} (\bibinfo {year} {2012})},\ \Eprint {http://arxiv.org/abs/1206.5712}
  {arXiv:1206.5712 [astro-ph.SR]} \BibitemShut {NoStop}%
\bibitem [{\citenamefont {{Thurgood}}\ and\ \citenamefont
  {{Tsiklauri}}(2015)}]{2015A&A...584A..83T}%
  \BibitemOpen
  \bibfield  {author} {\bibinfo {author} {\bibfnamefont {J.~O.}\ \bibnamefont
  {{Thurgood}}}\ and\ \bibinfo {author} {\bibfnamefont {D.}~\bibnamefont
  {{Tsiklauri}}},\ }\href {\doibase 10.1051/0004-6361/201527079} {\bibfield
  {journal} {\bibinfo  {journal} {Astron. Astrophys.}\ }\textbf {\bibinfo
  {volume} {584}},\ \bibinfo {eid} {A83} (\bibinfo {year} {2015})},\ \Eprint
  {http://arxiv.org/abs/1509.07004} {arXiv:1509.07004 [astro-ph.SR]}
  \BibitemShut {NoStop}%
\bibitem [{\citenamefont {{Henri}}\ \emph {et~al.}(2019)\citenamefont
  {{Henri}}, \citenamefont {{Sgattoni}}, \citenamefont {{Briand}},
  \citenamefont {{Amiranoff}},\ and\ \citenamefont
  {{Riconda}}}]{2019JGRA..124.1475H}%
  \BibitemOpen
  \bibfield  {author} {\bibinfo {author} {\bibfnamefont {P.}~\bibnamefont
  {{Henri}}}, \bibinfo {author} {\bibfnamefont {A.}~\bibnamefont {{Sgattoni}}},
  \bibinfo {author} {\bibfnamefont {C.}~\bibnamefont {{Briand}}}, \bibinfo
  {author} {\bibfnamefont {F.}~\bibnamefont {{Amiranoff}}}, \ and\ \bibinfo
  {author} {\bibfnamefont {C.}~\bibnamefont {{Riconda}}},\ }\href {\doibase
  10.1029/2018JA025707} {\bibfield  {journal} {\bibinfo  {journal} {Journal of
  Geophysical Research (Space Physics)}\ }\textbf {\bibinfo {volume} {124}},\
  \bibinfo {pages} {1475} (\bibinfo {year} {2019})}\BibitemShut {NoStop}%
\bibitem [{\citenamefont {{Lee}}\ \emph {et~al.}(2019)\citenamefont {{Lee}},
  \citenamefont {{Ziebell}}, \citenamefont {{Yoon}}, \citenamefont
  {{Gaelzer}},\ and\ \citenamefont {{Lee}}}]{2019ApJ...871...74L}%
  \BibitemOpen
  \bibfield  {author} {\bibinfo {author} {\bibfnamefont {S.-Y.}\ \bibnamefont
  {{Lee}}}, \bibinfo {author} {\bibfnamefont {L.~F.}\ \bibnamefont
  {{Ziebell}}}, \bibinfo {author} {\bibfnamefont {P.~H.}\ \bibnamefont
  {{Yoon}}}, \bibinfo {author} {\bibfnamefont {R.}~\bibnamefont {{Gaelzer}}}, \
  and\ \bibinfo {author} {\bibfnamefont {E.~S.}\ \bibnamefont {{Lee}}},\ }\href
  {\doibase 10.3847/1538-4357/aaf476} {\bibfield  {journal} {\bibinfo
  {journal} {\apj}\ }\textbf {\bibinfo {volume} {871}},\ \bibinfo {eid} {74}
  (\bibinfo {year} {2019})},\ \Eprint {http://arxiv.org/abs/1811.02392}
  {arXiv:1811.02392 [astro-ph.SR]} \BibitemShut {NoStop}%
\bibitem [{\citenamefont {{Krafft}}\ and\ \citenamefont
  {{Volokitin}}(2021)}]{2021ApJ...923..103K}%
  \BibitemOpen
  \bibfield  {author} {\bibinfo {author} {\bibfnamefont {C.}~\bibnamefont
  {{Krafft}}}\ and\ \bibinfo {author} {\bibfnamefont {A.~S.}\ \bibnamefont
  {{Volokitin}}},\ }\href {\doibase 10.3847/1538-4357/ac2153} {\bibfield
  {journal} {\bibinfo  {journal} {\apj}\ }\textbf {\bibinfo {volume} {923}},\
  \bibinfo {eid} {103} (\bibinfo {year} {2021})}\BibitemShut {NoStop}%
\bibitem [{\citenamefont {{Krafft}}\ and\ \citenamefont
  {{Savoini}}(2022)}]{2022ApJ...924L..24K}%
  \BibitemOpen
  \bibfield  {author} {\bibinfo {author} {\bibfnamefont {C.}~\bibnamefont
  {{Krafft}}}\ and\ \bibinfo {author} {\bibfnamefont {P.}~\bibnamefont
  {{Savoini}}},\ }\href {\doibase 10.3847/2041-8213/ac46a7} {\bibfield
  {journal} {\bibinfo  {journal} {Astrophys. J. Lett.}\ }\textbf {\bibinfo
  {volume} {924}},\ \bibinfo {eid} {L24} (\bibinfo {year} {2022})}\BibitemShut
  {NoStop}%
\bibitem [{\citenamefont {{Wu}}\ \emph {et~al.}(1989)\citenamefont {{Wu}},
  \citenamefont {{Lui}}, \citenamefont {{Mandt}},\ and\ \citenamefont
  {{Krauss-Varban}}}]{1989GeoRL..16.1125W}%
  \BibitemOpen
  \bibfield  {author} {\bibinfo {author} {\bibfnamefont {C.~S.}\ \bibnamefont
  {{Wu}}}, \bibinfo {author} {\bibfnamefont {A.~T.~Y.}\ \bibnamefont {{Lui}}},
  \bibinfo {author} {\bibfnamefont {M.~E.}\ \bibnamefont {{Mandt}}}, \ and\
  \bibinfo {author} {\bibfnamefont {D.}~\bibnamefont {{Krauss-Varban}}},\
  }\href {\doibase 10.1029/GL016i010p01125} {\bibfield  {journal} {\bibinfo
  {journal} {Geophysical Research Letters}\ }\textbf {\bibinfo {volume} {16}},\
  \bibinfo {pages} {1125} (\bibinfo {year} {1989})}\BibitemShut {NoStop}%
\bibitem [{\citenamefont {{Leroy}}\ and\ \citenamefont
  {{Mangeney}}(1984)}]{1984AnGeo...2..449L}%
  \BibitemOpen
  \bibfield  {author} {\bibinfo {author} {\bibfnamefont {M.~M.}\ \bibnamefont
  {{Leroy}}}\ and\ \bibinfo {author} {\bibfnamefont {A.}~\bibnamefont
  {{Mangeney}}},\ }\href@noop {} {\bibfield  {journal} {\bibinfo  {journal}
  {Annales Geophysicae}\ }\textbf {\bibinfo {volume} {2}},\ \bibinfo {pages}
  {449} (\bibinfo {year} {1984})}\BibitemShut {NoStop}%
\bibitem [{\citenamefont {{Vlahos}}\ and\ \citenamefont
  {{Sprangle}}(1987)}]{1987ApJ...322..463V}%
  \BibitemOpen
  \bibfield  {author} {\bibinfo {author} {\bibfnamefont {L.}~\bibnamefont
  {{Vlahos}}}\ and\ \bibinfo {author} {\bibfnamefont {P.}~\bibnamefont
  {{Sprangle}}},\ }\href {\doibase 10.1086/165742} {\bibfield  {journal}
  {\bibinfo  {journal} {\apj}\ }\textbf {\bibinfo {volume} {322}},\ \bibinfo
  {pages} {463} (\bibinfo {year} {1987})}\BibitemShut {NoStop}%
\bibitem [{\citenamefont {{Vlahos}}(1987)}]{1987SoPh..111..155V}%
  \BibitemOpen
  \bibfield  {author} {\bibinfo {author} {\bibfnamefont {L.}~\bibnamefont
  {{Vlahos}}},\ }\href {\doibase 10.1007/BF00145448} {\bibfield  {journal}
  {\bibinfo  {journal} {Sol. Phys.}\ }\textbf {\bibinfo {volume} {111}},\
  \bibinfo {pages} {155} (\bibinfo {year} {1987})}\BibitemShut {NoStop}%
\bibitem [{\citenamefont {{Yang}}\ \emph {et~al.}(2020)\citenamefont {{Yang}},
  \citenamefont {{Liu}}, \citenamefont {{Matsukiyo}}, \citenamefont {{Lu}},
  \citenamefont {{Guo}}, \citenamefont {{Liu}}, \citenamefont {{Xie}},
  \citenamefont {{Gao}},\ and\ \citenamefont {{Guo}}}]{2020ApJ...900L..24Y}%
  \BibitemOpen
  \bibfield  {author} {\bibinfo {author} {\bibfnamefont {Z.}~\bibnamefont
  {{Yang}}}, \bibinfo {author} {\bibfnamefont {Y.~D.}\ \bibnamefont {{Liu}}},
  \bibinfo {author} {\bibfnamefont {S.}~\bibnamefont {{Matsukiyo}}}, \bibinfo
  {author} {\bibfnamefont {Q.}~\bibnamefont {{Lu}}}, \bibinfo {author}
  {\bibfnamefont {F.}~\bibnamefont {{Guo}}}, \bibinfo {author} {\bibfnamefont
  {M.}~\bibnamefont {{Liu}}}, \bibinfo {author} {\bibfnamefont
  {H.}~\bibnamefont {{Xie}}}, \bibinfo {author} {\bibfnamefont
  {X.}~\bibnamefont {{Gao}}}, \ and\ \bibinfo {author} {\bibfnamefont
  {J.}~\bibnamefont {{Guo}}},\ }\href {\doibase 10.3847/2041-8213/abaf59}
  {\bibfield  {journal} {\bibinfo  {journal} {Astrophys. J. Lett.}\ }\textbf
  {\bibinfo {volume} {900}},\ \bibinfo {eid} {L24} (\bibinfo {year} {2020})},\
  \Eprint {http://arxiv.org/abs/2008.06820} {arXiv:2008.06820
  [physics.space-ph]} \BibitemShut {NoStop}%
\bibitem [{\citenamefont {{Bessho}}\ \emph {et~al.}(2014)\citenamefont
  {{Bessho}}, \citenamefont {{Chen}}, \citenamefont {{Shuster}},\ and\
  \citenamefont {{Wang}}}]{2014GeoRL..41.8688B}%
  \BibitemOpen
  \bibfield  {author} {\bibinfo {author} {\bibfnamefont {N.}~\bibnamefont
  {{Bessho}}}, \bibinfo {author} {\bibfnamefont {L.~J.}\ \bibnamefont
  {{Chen}}}, \bibinfo {author} {\bibfnamefont {J.~R.}\ \bibnamefont
  {{Shuster}}}, \ and\ \bibinfo {author} {\bibfnamefont {S.}~\bibnamefont
  {{Wang}}},\ }\href {\doibase 10.1002/2014GL062034} {\bibfield  {journal}
  {\bibinfo  {journal} {Geophysical Research Letters}\ }\textbf {\bibinfo
  {volume} {41}},\ \bibinfo {pages} {8688} (\bibinfo {year}
  {2014})}\BibitemShut {NoStop}%
\bibitem [{\citenamefont {{Shuster}}\ \emph {et~al.}(2014)\citenamefont
  {{Shuster}}, \citenamefont {{Chen}}, \citenamefont {{Daughton}},
  \citenamefont {{Lee}}, \citenamefont {{Lee}}, \citenamefont {{Bessho}},
  \citenamefont {{Torbert}}, \citenamefont {{Li}},\ and\ \citenamefont
  {{Argall}}}]{2014GeoRL..41.5389S}%
  \BibitemOpen
  \bibfield  {author} {\bibinfo {author} {\bibfnamefont {J.~R.}\ \bibnamefont
  {{Shuster}}}, \bibinfo {author} {\bibfnamefont {L.~J.}\ \bibnamefont
  {{Chen}}}, \bibinfo {author} {\bibfnamefont {W.~S.}\ \bibnamefont
  {{Daughton}}}, \bibinfo {author} {\bibfnamefont {L.~C.}\ \bibnamefont
  {{Lee}}}, \bibinfo {author} {\bibfnamefont {K.~H.}\ \bibnamefont {{Lee}}},
  \bibinfo {author} {\bibfnamefont {N.}~\bibnamefont {{Bessho}}}, \bibinfo
  {author} {\bibfnamefont {R.~B.}\ \bibnamefont {{Torbert}}}, \bibinfo {author}
  {\bibfnamefont {G.}~\bibnamefont {{Li}}}, \ and\ \bibinfo {author}
  {\bibfnamefont {M.~R.}\ \bibnamefont {{Argall}}},\ }\href {\doibase
  10.1002/2014GL060608} {\bibfield  {journal} {\bibinfo  {journal} {Geophysical
  Research Letters}\ }\textbf {\bibinfo {volume} {41}},\ \bibinfo {pages}
  {5389} (\bibinfo {year} {2014})}\BibitemShut {NoStop}%
\bibitem [{\citenamefont {{Shuster}}\ \emph {et~al.}(2015)\citenamefont
  {{Shuster}}, \citenamefont {{Chen}}, \citenamefont {{Hesse}}, \citenamefont
  {{Argall}}, \citenamefont {{Daughton}}, \citenamefont {{Torbert}},\ and\
  \citenamefont {{Bessho}}}]{2015GeoRL..42.2586S}%
  \BibitemOpen
  \bibfield  {author} {\bibinfo {author} {\bibfnamefont {J.~R.}\ \bibnamefont
  {{Shuster}}}, \bibinfo {author} {\bibfnamefont {L.~J.}\ \bibnamefont
  {{Chen}}}, \bibinfo {author} {\bibfnamefont {M.}~\bibnamefont {{Hesse}}},
  \bibinfo {author} {\bibfnamefont {M.~R.}\ \bibnamefont {{Argall}}}, \bibinfo
  {author} {\bibfnamefont {W.}~\bibnamefont {{Daughton}}}, \bibinfo {author}
  {\bibfnamefont {R.~B.}\ \bibnamefont {{Torbert}}}, \ and\ \bibinfo {author}
  {\bibfnamefont {N.}~\bibnamefont {{Bessho}}},\ }\href {\doibase
  10.1002/2015GL063601} {\bibfield  {journal} {\bibinfo  {journal} {Geophysical
  Research Letters}\ }\textbf {\bibinfo {volume} {42}},\ \bibinfo {pages}
  {2586} (\bibinfo {year} {2015})}\BibitemShut {NoStop}%
\bibitem [{\citenamefont {{Li}}\ \emph {et~al.}(2017)\citenamefont {{Li}},
  \citenamefont {{Chen}}, \citenamefont {{Wang}}, \citenamefont {{Ruan}},
  \citenamefont {{Feng}}, \citenamefont {{Du}},\ and\ \citenamefont
  {{Kong}}}]{2017SoPh..292...82L}%
  \BibitemOpen
  \bibfield  {author} {\bibinfo {author} {\bibfnamefont {C.~Y.}\ \bibnamefont
  {{Li}}}, \bibinfo {author} {\bibfnamefont {Y.}~\bibnamefont {{Chen}}},
  \bibinfo {author} {\bibfnamefont {B.}~\bibnamefont {{Wang}}}, \bibinfo
  {author} {\bibfnamefont {G.~P.}\ \bibnamefont {{Ruan}}}, \bibinfo {author}
  {\bibfnamefont {S.~W.}\ \bibnamefont {{Feng}}}, \bibinfo {author}
  {\bibfnamefont {G.~H.}\ \bibnamefont {{Du}}}, \ and\ \bibinfo {author}
  {\bibfnamefont {X.~L.}\ \bibnamefont {{Kong}}},\ }\href {\doibase
  10.1007/s11207-017-1108-1} {\bibfield  {journal} {\bibinfo  {journal} {Sol.
  Phys.}\ }\textbf {\bibinfo {volume} {292}},\ \bibinfo {eid} {82} (\bibinfo
  {year} {2017})},\ \Eprint {http://arxiv.org/abs/1705.01666} {arXiv:1705.01666
  [astro-ph.SR]} \BibitemShut {NoStop}%
\bibitem [{\citenamefont {{Feng}}\ \emph {et~al.}(2012)\citenamefont {{Feng}},
  \citenamefont {{Chen}}, \citenamefont {{Kong}}, \citenamefont {{Li}},
  \citenamefont {{Song}}, \citenamefont {{Feng}},\ and\ \citenamefont
  {{Liu}}}]{2012ApJ...753...21F}%
  \BibitemOpen
  \bibfield  {author} {\bibinfo {author} {\bibfnamefont {S.~W.}\ \bibnamefont
  {{Feng}}}, \bibinfo {author} {\bibfnamefont {Y.}~\bibnamefont {{Chen}}},
  \bibinfo {author} {\bibfnamefont {X.~L.}\ \bibnamefont {{Kong}}}, \bibinfo
  {author} {\bibfnamefont {G.}~\bibnamefont {{Li}}}, \bibinfo {author}
  {\bibfnamefont {H.~Q.}\ \bibnamefont {{Song}}}, \bibinfo {author}
  {\bibfnamefont {X.~S.}\ \bibnamefont {{Feng}}}, \ and\ \bibinfo {author}
  {\bibfnamefont {Y.}~\bibnamefont {{Liu}}},\ }\href {\doibase
  10.1088/0004-637X/753/1/21} {\bibfield  {journal} {\bibinfo  {journal}
  {\apj}\ }\textbf {\bibinfo {volume} {753}},\ \bibinfo {eid} {21} (\bibinfo
  {year} {2012})},\ \Eprint {http://arxiv.org/abs/1204.5569} {arXiv:1204.5569
  [astro-ph.SR]} \BibitemShut {NoStop}%
\bibitem [{\citenamefont {{Kong}}\ \emph {et~al.}(2012)\citenamefont {{Kong}},
  \citenamefont {{Chen}}, \citenamefont {{Li}}, \citenamefont {{Feng}},
  \citenamefont {{Song}}, \citenamefont {{Guo}},\ and\ \citenamefont
  {{Jiao}}}]{2012ApJ...750..158K}%
  \BibitemOpen
  \bibfield  {author} {\bibinfo {author} {\bibfnamefont {X.~L.}\ \bibnamefont
  {{Kong}}}, \bibinfo {author} {\bibfnamefont {Y.}~\bibnamefont {{Chen}}},
  \bibinfo {author} {\bibfnamefont {G.}~\bibnamefont {{Li}}}, \bibinfo {author}
  {\bibfnamefont {S.~W.}\ \bibnamefont {{Feng}}}, \bibinfo {author}
  {\bibfnamefont {H.~Q.}\ \bibnamefont {{Song}}}, \bibinfo {author}
  {\bibfnamefont {F.}~\bibnamefont {{Guo}}}, \ and\ \bibinfo {author}
  {\bibfnamefont {F.~R.}\ \bibnamefont {{Jiao}}},\ }\href {\doibase
  10.1088/0004-637X/750/2/158} {\bibfield  {journal} {\bibinfo  {journal}
  {\apj}\ }\textbf {\bibinfo {volume} {750}},\ \bibinfo {eid} {158} (\bibinfo
  {year} {2012})},\ \Eprint {http://arxiv.org/abs/1203.1511} {arXiv:1203.1511
  [astro-ph.SR]} \BibitemShut {NoStop}%
\bibitem [{\citenamefont {{Chen}}\ \emph {et~al.}(2014)\citenamefont {{Chen}},
  \citenamefont {{Du}}, \citenamefont {{Feng}}, \citenamefont {{Feng}},
  \citenamefont {{Kong}}, \citenamefont {{Guo}}, \citenamefont {{Wang}},\ and\
  \citenamefont {{Li}}}]{2014ApJ...787...59C}%
  \BibitemOpen
  \bibfield  {author} {\bibinfo {author} {\bibfnamefont {Y.}~\bibnamefont
  {{Chen}}}, \bibinfo {author} {\bibfnamefont {G.}~\bibnamefont {{Du}}},
  \bibinfo {author} {\bibfnamefont {L.}~\bibnamefont {{Feng}}}, \bibinfo
  {author} {\bibfnamefont {S.}~\bibnamefont {{Feng}}}, \bibinfo {author}
  {\bibfnamefont {X.}~\bibnamefont {{Kong}}}, \bibinfo {author} {\bibfnamefont
  {F.}~\bibnamefont {{Guo}}}, \bibinfo {author} {\bibfnamefont
  {B.}~\bibnamefont {{Wang}}}, \ and\ \bibinfo {author} {\bibfnamefont
  {G.}~\bibnamefont {{Li}}},\ }\href {\doibase 10.1088/0004-637X/787/1/59}
  {\bibfield  {journal} {\bibinfo  {journal} {\apj}\ }\textbf {\bibinfo
  {volume} {787}},\ \bibinfo {eid} {59} (\bibinfo {year} {2014})},\ \Eprint
  {http://arxiv.org/abs/1404.3052} {arXiv:1404.3052 [physics.space-ph]}
  \BibitemShut {NoStop}%
\bibitem [{\citenamefont {{Vasanth}}\ \emph {et~al.}(2016)\citenamefont
  {{Vasanth}}, \citenamefont {{Chen}}, \citenamefont {{Feng}}, \citenamefont
  {{Ma}}, \citenamefont {{Du}}, \citenamefont {{Song}}, \citenamefont
  {{Kong}},\ and\ \citenamefont {{Wang}}}]{2016ApJ...830L...2V}%
  \BibitemOpen
  \bibfield  {author} {\bibinfo {author} {\bibfnamefont {V.}~\bibnamefont
  {{Vasanth}}}, \bibinfo {author} {\bibfnamefont {Y.}~\bibnamefont {{Chen}}},
  \bibinfo {author} {\bibfnamefont {S.}~\bibnamefont {{Feng}}}, \bibinfo
  {author} {\bibfnamefont {S.}~\bibnamefont {{Ma}}}, \bibinfo {author}
  {\bibfnamefont {G.}~\bibnamefont {{Du}}}, \bibinfo {author} {\bibfnamefont
  {H.}~\bibnamefont {{Song}}}, \bibinfo {author} {\bibfnamefont
  {X.}~\bibnamefont {{Kong}}}, \ and\ \bibinfo {author} {\bibfnamefont
  {B.}~\bibnamefont {{Wang}}},\ }\href {\doibase 10.3847/2041-8205/830/1/L2}
  {\bibfield  {journal} {\bibinfo  {journal} {Astrophys. J. Lett.}\ }\textbf
  {\bibinfo {volume} {830}},\ \bibinfo {eid} {L2} (\bibinfo {year} {2016})},\
  \Eprint {http://arxiv.org/abs/1609.06546} {arXiv:1609.06546 [astro-ph.SR]}
  \BibitemShut {NoStop}%
\bibitem [{\citenamefont {{Vasanth}}\ \emph {et~al.}(2019)\citenamefont
  {{Vasanth}}, \citenamefont {{Chen}}, \citenamefont {{Lv}}, \citenamefont
  {{Ning}}, \citenamefont {{Li}}, \citenamefont {{Feng}}, \citenamefont
  {{Wu}},\ and\ \citenamefont {{Du}}}]{2019ApJ...870...30V}%
  \BibitemOpen
  \bibfield  {author} {\bibinfo {author} {\bibfnamefont {V.}~\bibnamefont
  {{Vasanth}}}, \bibinfo {author} {\bibfnamefont {Y.}~\bibnamefont {{Chen}}},
  \bibinfo {author} {\bibfnamefont {M.}~\bibnamefont {{Lv}}}, \bibinfo {author}
  {\bibfnamefont {H.}~\bibnamefont {{Ning}}}, \bibinfo {author} {\bibfnamefont
  {C.}~\bibnamefont {{Li}}}, \bibinfo {author} {\bibfnamefont {S.}~\bibnamefont
  {{Feng}}}, \bibinfo {author} {\bibfnamefont {Z.}~\bibnamefont {{Wu}}}, \ and\
  \bibinfo {author} {\bibfnamefont {G.}~\bibnamefont {{Du}}},\ }\href {\doibase
  10.3847/1538-4357/aaeffd} {\bibfield  {journal} {\bibinfo  {journal} {\apj}\
  }\textbf {\bibinfo {volume} {870}},\ \bibinfo {eid} {30} (\bibinfo {year}
  {2019})},\ \Eprint {http://arxiv.org/abs/1810.11815} {arXiv:1810.11815
  [astro-ph.SR]} \BibitemShut {NoStop}%
\bibitem [{\citenamefont {{Ni}}\ \emph {et~al.}(2020)\citenamefont {{Ni}},
  \citenamefont {{Chen}}, \citenamefont {{Li}}, \citenamefont {{Zhang}},
  \citenamefont {{Ning}}, \citenamefont {{Kong}}, \citenamefont {{Wang}},\ and\
  \citenamefont {{Hosseinpour}}}]{2020ApJ...891L..25N}%
  \BibitemOpen
  \bibfield  {author} {\bibinfo {author} {\bibfnamefont {S.}~\bibnamefont
  {{Ni}}}, \bibinfo {author} {\bibfnamefont {Y.}~\bibnamefont {{Chen}}},
  \bibinfo {author} {\bibfnamefont {C.}~\bibnamefont {{Li}}}, \bibinfo {author}
  {\bibfnamefont {Z.}~\bibnamefont {{Zhang}}}, \bibinfo {author} {\bibfnamefont
  {H.}~\bibnamefont {{Ning}}}, \bibinfo {author} {\bibfnamefont
  {X.}~\bibnamefont {{Kong}}}, \bibinfo {author} {\bibfnamefont
  {B.}~\bibnamefont {{Wang}}}, \ and\ \bibinfo {author} {\bibfnamefont
  {M.}~\bibnamefont {{Hosseinpour}}},\ }\href {\doibase
  10.3847/2041-8213/ab7750} {\bibfield  {journal} {\bibinfo  {journal}
  {Astrophys. J. Lett.}\ }\textbf {\bibinfo {volume} {891}},\ \bibinfo {eid}
  {L25} (\bibinfo {year} {2020})}\BibitemShut {NoStop}%
\bibitem [{\citenamefont {{Dory}}, \citenamefont {{Guest}},\ and\ \citenamefont
  {{Harris}}(1965)}]{1965PhRvL..14..131D}%
  \BibitemOpen
  \bibfield  {author} {\bibinfo {author} {\bibfnamefont {R.~A.}\ \bibnamefont
  {{Dory}}}, \bibinfo {author} {\bibfnamefont {G.~E.}\ \bibnamefont {{Guest}}},
  \ and\ \bibinfo {author} {\bibfnamefont {E.~G.}\ \bibnamefont {{Harris}}},\
  }\href {\doibase 10.1103/PhysRevLett.14.131} {\bibfield  {journal} {\bibinfo
  {journal} {\prl}\ }\textbf {\bibinfo {volume} {14}},\ \bibinfo {pages} {131}
  (\bibinfo {year} {1965})}\BibitemShut {NoStop}%
\bibitem [{\citenamefont {{Wu}}\ and\ \citenamefont
  {{Lee}}(1979)}]{1979ApJ...230..621W}%
  \BibitemOpen
  \bibfield  {author} {\bibinfo {author} {\bibfnamefont {C.~S.}\ \bibnamefont
  {{Wu}}}\ and\ \bibinfo {author} {\bibfnamefont {L.~C.}\ \bibnamefont
  {{Lee}}},\ }\href {\doibase 10.1086/157120} {\bibfield  {journal} {\bibinfo
  {journal} {\apj}\ }\textbf {\bibinfo {volume} {230}},\ \bibinfo {pages} {621}
  (\bibinfo {year} {1979})}\BibitemShut {NoStop}%
\bibitem [{\citenamefont {{Lee}}\ \emph {et~al.}(2009)\citenamefont {{Lee}},
  \citenamefont {{Omura}}, \citenamefont {{Lee}},\ and\ \citenamefont
  {{Wu}}}]{2009PhRvL.103j5101L}%
  \BibitemOpen
  \bibfield  {author} {\bibinfo {author} {\bibfnamefont {K.~H.}\ \bibnamefont
  {{Lee}}}, \bibinfo {author} {\bibfnamefont {Y.}~\bibnamefont {{Omura}}},
  \bibinfo {author} {\bibfnamefont {L.~C.}\ \bibnamefont {{Lee}}}, \ and\
  \bibinfo {author} {\bibfnamefont {C.~S.}\ \bibnamefont {{Wu}}},\ }\href
  {\doibase 10.1103/PhysRevLett.103.105101} {\bibfield  {journal} {\bibinfo
  {journal} {\prl}\ }\textbf {\bibinfo {volume} {103}},\ \bibinfo {eid}
  {105101} (\bibinfo {year} {2009})}\BibitemShut {NoStop}%
\bibitem [{\citenamefont {{Lee}}, \citenamefont {{Omura}},\ and\ \citenamefont
  {{Lee}}(2011)}]{2011PhPl...18i2110L}%
  \BibitemOpen
  \bibfield  {author} {\bibinfo {author} {\bibfnamefont {K.~H.}\ \bibnamefont
  {{Lee}}}, \bibinfo {author} {\bibfnamefont {Y.}~\bibnamefont {{Omura}}}, \
  and\ \bibinfo {author} {\bibfnamefont {L.~C.}\ \bibnamefont {{Lee}}},\ }\href
  {\doibase 10.1063/1.3626562} {\bibfield  {journal} {\bibinfo  {journal}
  {Physics of Plasmas}\ }\textbf {\bibinfo {volume} {18}},\ \bibinfo {eid}
  {092110} (\bibinfo {year} {2011})}\BibitemShut {NoStop}%
\bibitem [{\citenamefont {{Zhou}}\ \emph {et~al.}(2020)\citenamefont {{Zhou}},
  \citenamefont {{Mu{\~n}oz}}, \citenamefont {{B{\"u}chner}},\ and\
  \citenamefont {{Liu}}}]{2020ApJ...891...92Z}%
  \BibitemOpen
  \bibfield  {author} {\bibinfo {author} {\bibfnamefont {X.}~\bibnamefont
  {{Zhou}}}, \bibinfo {author} {\bibfnamefont {P.~A.}\ \bibnamefont
  {{Mu{\~n}oz}}}, \bibinfo {author} {\bibfnamefont {J.}~\bibnamefont
  {{B{\"u}chner}}}, \ and\ \bibinfo {author} {\bibfnamefont {S.}~\bibnamefont
  {{Liu}}},\ }\href {\doibase 10.3847/1538-4357/ab6a0d} {\bibfield  {journal}
  {\bibinfo  {journal} {\apj}\ }\textbf {\bibinfo {volume} {891}},\ \bibinfo
  {eid} {92} (\bibinfo {year} {2020})},\ \Eprint
  {http://arxiv.org/abs/1907.12958} {arXiv:1907.12958 [physics.plasm-ph]}
  \BibitemShut {NoStop}%
\bibitem [{\citenamefont {{Chen}}\ \emph {et~al.}(2022)\citenamefont {{Chen}},
  \citenamefont {{Zhang}}, \citenamefont {{Ni}}, \citenamefont {{Li}},
  \citenamefont {{Ning}},\ and\ \citenamefont {{Kong}}}]{2022ApJ...924L..34C}%
  \BibitemOpen
  \bibfield  {author} {\bibinfo {author} {\bibfnamefont {Y.}~\bibnamefont
  {{Chen}}}, \bibinfo {author} {\bibfnamefont {Z.}~\bibnamefont {{Zhang}}},
  \bibinfo {author} {\bibfnamefont {S.}~\bibnamefont {{Ni}}}, \bibinfo {author}
  {\bibfnamefont {C.}~\bibnamefont {{Li}}}, \bibinfo {author} {\bibfnamefont
  {H.}~\bibnamefont {{Ning}}}, \ and\ \bibinfo {author} {\bibfnamefont
  {X.}~\bibnamefont {{Kong}}},\ }\href {\doibase 10.3847/2041-8213/ac47fa}
  {\bibfield  {journal} {\bibinfo  {journal} {Astrophys. J. Lett.}\ }\textbf
  {\bibinfo {volume} {924}},\ \bibinfo {eid} {L34} (\bibinfo {year} {2022})},\
  \Eprint {http://arxiv.org/abs/2201.03937} {arXiv:2201.03937
  [physics.plasm-ph]} \BibitemShut {NoStop}%
\bibitem [{\citenamefont {{Cairns}}(1989)}]{1989PhFlB...1..204C}%
  \BibitemOpen
  \bibfield  {author} {\bibinfo {author} {\bibfnamefont {I.~H.}\ \bibnamefont
  {{Cairns}}},\ }\href {\doibase 10.1063/1.859088} {\bibfield  {journal}
  {\bibinfo  {journal} {Physics of Fluids B}\ }\textbf {\bibinfo {volume}
  {1}},\ \bibinfo {pages} {204} (\bibinfo {year} {1989})}\BibitemShut {NoStop}%
\bibitem [{\citenamefont {{Zhang}}\ \emph {et~al.}(2022)\citenamefont
  {{Zhang}}, \citenamefont {{Chen}}, \citenamefont {{Ni}}, \citenamefont
  {{Li}}, \citenamefont {{Ning}}, \citenamefont {{Li}},\ and\ \citenamefont
  {{Kong}}}]{2022arXiv220911707Z}%
  \BibitemOpen
  \bibfield  {author} {\bibinfo {author} {\bibfnamefont {Z.}~\bibnamefont
  {{Zhang}}}, \bibinfo {author} {\bibfnamefont {Y.}~\bibnamefont {{Chen}}},
  \bibinfo {author} {\bibfnamefont {S.}~\bibnamefont {{Ni}}}, \bibinfo {author}
  {\bibfnamefont {C.}~\bibnamefont {{Li}}}, \bibinfo {author} {\bibfnamefont
  {H.}~\bibnamefont {{Ning}}}, \bibinfo {author} {\bibfnamefont
  {Y.}~\bibnamefont {{Li}}}, \ and\ \bibinfo {author} {\bibfnamefont
  {X.}~\bibnamefont {{Kong}}},\ }\href@noop {} {\bibfield  {journal} {\bibinfo
  {journal} {arXiv e-prints}\ ,\ \bibinfo {eid} {arXiv:2209.11707}} (\bibinfo
  {year} {2022})},\ \Eprint {http://arxiv.org/abs/2209.11707} {arXiv:2209.11707
  [physics.plasm-ph]} \BibitemShut {NoStop}%
\bibitem [{\citenamefont {{Bowers}}\ \emph {et~al.}(2008)\citenamefont
  {{Bowers}}, \citenamefont {{Albright}}, \citenamefont {{Yin}}, \citenamefont
  {{Bergen}},\ and\ \citenamefont {{Kwan}}}]{2008PhPl...15e5703B}%
  \BibitemOpen
  \bibfield  {author} {\bibinfo {author} {\bibfnamefont {K.~J.}\ \bibnamefont
  {{Bowers}}}, \bibinfo {author} {\bibfnamefont {B.~J.}\ \bibnamefont
  {{Albright}}}, \bibinfo {author} {\bibfnamefont {L.}~\bibnamefont {{Yin}}},
  \bibinfo {author} {\bibfnamefont {B.}~\bibnamefont {{Bergen}}}, \ and\
  \bibinfo {author} {\bibfnamefont {T.~J.~T.}\ \bibnamefont {{Kwan}}},\ }\href
  {\doibase 10.1063/1.2840133} {\bibfield  {journal} {\bibinfo  {journal}
  {Physics of Plasmas}\ }\textbf {\bibinfo {volume} {15}},\ \bibinfo {eid}
  {055703} (\bibinfo {year} {2008})}\BibitemShut {NoStop}%
\bibitem [{\citenamefont {Bowers}\ \emph {et~al.}(2008)\citenamefont {Bowers},
  \citenamefont {Albright}, \citenamefont {Bergen}, \citenamefont {Yin},
  \citenamefont {Barker},\ and\ \citenamefont {Kerbyson}}]{bowers20080}%
  \BibitemOpen
  \bibfield  {author} {\bibinfo {author} {\bibfnamefont {K.~J.}\ \bibnamefont
  {Bowers}}, \bibinfo {author} {\bibfnamefont {B.~J.}\ \bibnamefont
  {Albright}}, \bibinfo {author} {\bibfnamefont {B.}~\bibnamefont {Bergen}},
  \bibinfo {author} {\bibfnamefont {L.}~\bibnamefont {Yin}}, \bibinfo {author}
  {\bibfnamefont {K.~J.}\ \bibnamefont {Barker}}, \ and\ \bibinfo {author}
  {\bibfnamefont {D.~J.}\ \bibnamefont {Kerbyson}},\ }in\ \href@noop {} {\emph
  {\bibinfo {booktitle} {SC'08: Proceedings of the 2008 ACM/IEEE conference on
  Supercomputing}}}\ (\bibinfo {organization} {IEEE},\ \bibinfo {year} {2008})\
  pp.\ \bibinfo {pages} {1--11}\BibitemShut {NoStop}%
\bibitem [{\citenamefont {{Bowers}}\ \emph {et~al.}(2009)\citenamefont
  {{Bowers}}, \citenamefont {{Albright}}, \citenamefont {{Yin}}, \citenamefont
  {{Daughton}}, \citenamefont {{Roytershteyn}}, \citenamefont {{Bergen}},\ and\
  \citenamefont {{Kwan}}}]{2009JPhCS.180a2055B}%
  \BibitemOpen
  \bibfield  {author} {\bibinfo {author} {\bibfnamefont {K.~J.}\ \bibnamefont
  {{Bowers}}}, \bibinfo {author} {\bibfnamefont {B.~J.}\ \bibnamefont
  {{Albright}}}, \bibinfo {author} {\bibfnamefont {L.}~\bibnamefont {{Yin}}},
  \bibinfo {author} {\bibfnamefont {W.}~\bibnamefont {{Daughton}}}, \bibinfo
  {author} {\bibfnamefont {V.}~\bibnamefont {{Roytershteyn}}}, \bibinfo
  {author} {\bibfnamefont {B.}~\bibnamefont {{Bergen}}}, \ and\ \bibinfo
  {author} {\bibfnamefont {T.~J.~T.}\ \bibnamefont {{Kwan}}},\ }in\ \href
  {\doibase 10.1088/1742-6596/180/1/012055} {\emph {\bibinfo {booktitle}
  {Journal of Physics Conference Series}}},\ \bibinfo {series} {Journal of
  Physics Conference Series}, Vol.\ \bibinfo {volume} {180}\ (\bibinfo {year}
  {2009})\ p.\ \bibinfo {pages} {012055}\BibitemShut {NoStop}%
\bibitem [{\citenamefont {{Umeda}}\ \emph {et~al.}(2012)\citenamefont
  {{Umeda}}, \citenamefont {{Matsukiyo}}, \citenamefont {{Amano}},\ and\
  \citenamefont {{Miyoshi}}}]{2012PhPl...19g2107U}%
  \BibitemOpen
  \bibfield  {author} {\bibinfo {author} {\bibfnamefont {T.}~\bibnamefont
  {{Umeda}}}, \bibinfo {author} {\bibfnamefont {S.}~\bibnamefont
  {{Matsukiyo}}}, \bibinfo {author} {\bibfnamefont {T.}~\bibnamefont
  {{Amano}}}, \ and\ \bibinfo {author} {\bibfnamefont {Y.}~\bibnamefont
  {{Miyoshi}}},\ }\href {\doibase 10.1063/1.4736848} {\bibfield  {journal}
  {\bibinfo  {journal} {Physics of Plasmas}\ }\textbf {\bibinfo {volume}
  {19}},\ \bibinfo {eid} {072107} (\bibinfo {year} {2012})}\BibitemShut
  {NoStop}%
\bibitem [{\citenamefont {{Hadi}}, \citenamefont {{Yoon}},\ and\ \citenamefont
  {{Qamar}}(2015)}]{2015PhPl...22b2112H}%
  \BibitemOpen
  \bibfield  {author} {\bibinfo {author} {\bibfnamefont {F.}~\bibnamefont
  {{Hadi}}}, \bibinfo {author} {\bibfnamefont {P.~H.}\ \bibnamefont {{Yoon}}},
  \ and\ \bibinfo {author} {\bibfnamefont {A.}~\bibnamefont {{Qamar}}},\ }\href
  {\doibase 10.1063/1.4907657} {\bibfield  {journal} {\bibinfo  {journal}
  {Physics of Plasmas}\ }\textbf {\bibinfo {volume} {22}},\ \bibinfo {eid}
  {022112} (\bibinfo {year} {2015})}\BibitemShut {NoStop}%
\bibitem [{\citenamefont {{Davidson}}(1983)}]{1983bpp..conf..229D}%
  \BibitemOpen
  \bibfield  {author} {\bibinfo {author} {\bibfnamefont {R.~C.}\ \bibnamefont
  {{Davidson}}},\ }in\ \href@noop {} {\emph {\bibinfo {booktitle} {Basic Plasma
  Physics: Selected Chapters, Handbook of Plasma Physics, Volume 1}}}\
  (\bibinfo {year} {1983})\ p.\ \bibinfo {pages} {229}\BibitemShut {NoStop}%
\bibitem [{\citenamefont {{Li}}\ \emph {et~al.}(2021)\citenamefont {{Li}},
  \citenamefont {{Chen}}, \citenamefont {{Ni}}, \citenamefont {{Tan}},
  \citenamefont {{Ning}},\ and\ \citenamefont {{Zhang}}}]{2021ApJ...909L...5L}%
  \BibitemOpen
  \bibfield  {author} {\bibinfo {author} {\bibfnamefont {C.}~\bibnamefont
  {{Li}}}, \bibinfo {author} {\bibfnamefont {Y.}~\bibnamefont {{Chen}}},
  \bibinfo {author} {\bibfnamefont {S.}~\bibnamefont {{Ni}}}, \bibinfo {author}
  {\bibfnamefont {B.}~\bibnamefont {{Tan}}}, \bibinfo {author} {\bibfnamefont
  {H.}~\bibnamefont {{Ning}}}, \ and\ \bibinfo {author} {\bibfnamefont
  {Z.}~\bibnamefont {{Zhang}}},\ }\href {\doibase 10.3847/2041-8213/abe708}
  {\bibfield  {journal} {\bibinfo  {journal} {Astrophys. J. Lett.}\ }\textbf
  {\bibinfo {volume} {909}},\ \bibinfo {eid} {L5} (\bibinfo {year} {2021})},\
  \Eprint {http://arxiv.org/abs/2102.09172} {arXiv:2102.09172 [astro-ph.SR]}
  \BibitemShut {NoStop}%
\end{thebibliography}%

\begin{table}
\caption{Wave parameters ($\omega, k, \theta, n$) for the resonance curves plotted in Figure 2} 
\label{table:1} 
\centering 
\begin{tabular}{c c c c c c c c c c c c c c c c} 
\hline\hline 
  & \multicolumn{4}{c}{Panel (b)} & & \multicolumn{2}{c}{Panel (d)} & & \multicolumn{4}{c}{Panel (f)} & & \multicolumn{2}{c}{Panel (h)}\\ 
  \hline 
& UH1 & UH2 & Z & Z & & W$_\mathrm{H}$ & W$_\mathrm{L}$ & & UH1 & UH2 & Z & Z & & W$_\mathrm{H}$ & W$_\mathrm{L}$\\
\hline 
$\omega~(\wce$)& 9.9 & 9.9 & 9.84 & 9.72 & & 0.4 & 0.04 & & 9.9 & 9.9 & 9.84 & 9.72 & & 0.4 & 0.04\\
$k~(\wce/c)$   & 50 & 50   & 4.5  & 3.1  & & 8.2 & 2.95 & & 65 & 65 & 4.5 & 3.1 & & 8.2 & 2.95\\
$\theta$ & $90^{\degr}$ & $80^{\degr}$ & $80^{\degr}$ & $80^{\degr}$ & & $0^{\degr}$ & $60^{\degr}$ & & $90^{\degr}$ & $80^{\degr}$ & $80^{\degr}$ & $80^{\degr}$ & & $0^{\degr}$ & $60^{\degr}$\\
$n$ & 10 & 10 & 10 & 10 & & 1 & 1 & & 10 & 10 & 10 & 10 & & 1 & 1 \\
\hline 
\end{tabular}
\end{table}

\begin{figure}
\centering
\includegraphics[width=0.99\linewidth]{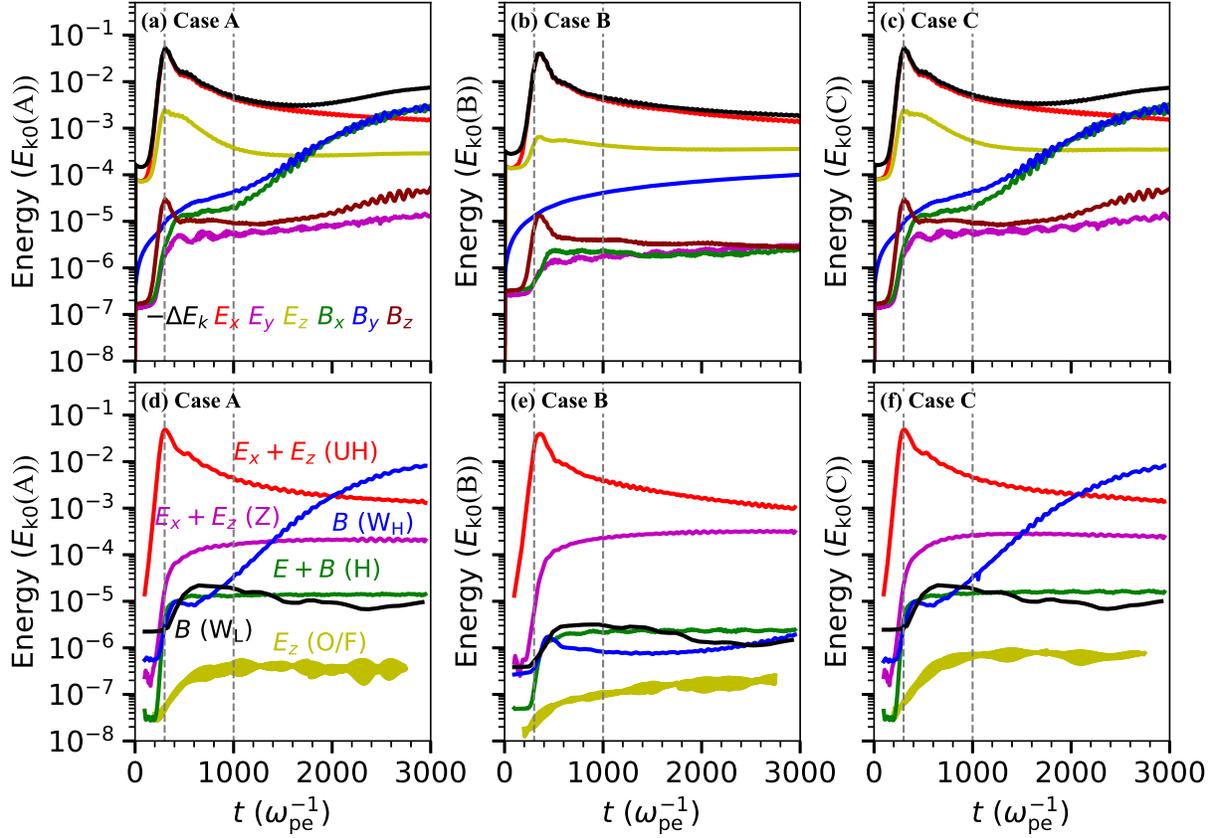}
  \caption{Upper panels: temporal profiles of energy of $E_x, E_y, E_z, B_x, B_y, B_z$ and -$\Delta E_k$
  (the negative change of the total kinetic energy of electrons) for Cases A--C;
  lower panels: temporal profiles of mode energy given by the dominating field
  components of each mode (UH for the upper-hybrid mode, Z for the slow extraordinary mode,
   W$_\mathrm{H}$ (W$_\mathrm{L}$) for the high- (low-) frequency component of the W mode, O/F for the fundamental emission in O mode and H for the harmonic
  emission, $B (E)$ represents the total magnetic- (electric-) field energy). The two vertical
  lines represent the ending times of Stage 1 (t = 300 $\wpe^{-1}$) and Stage 2 (t = 1000
  $\wpe^{-1}$).}
     \label{Fig1}
\end{figure}

\begin{figure}
\centering
\includegraphics[width=0.99\linewidth]{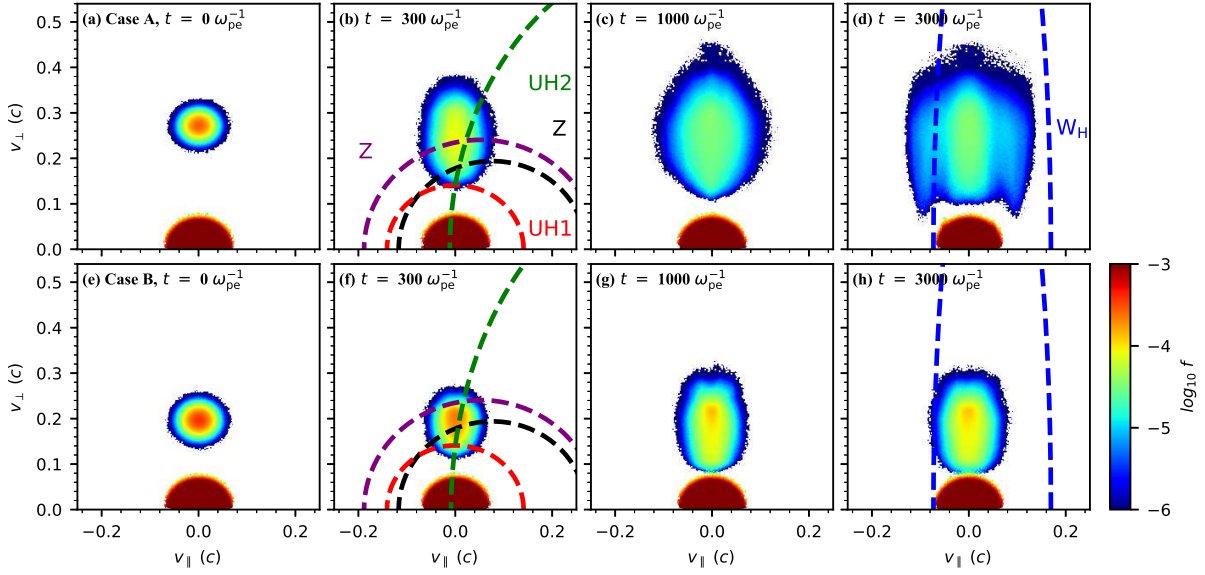}
  \caption{The electron VDFs at four moments for Case A (upper panels) and Case B (lower panels).
  Resonance curves are plotted in panel (b: Case A) and in panel (f: Case B) for UH1 (red-dashed),
  UH2 (green-dashed), Z (black-dashed and purple-dashed), and in panels d (Case A) and h (Case B) for W$_\mathrm{H}$ (blue-dashed). The resonance curves for W$_\mathrm{L}$ are too large in both coordinates to be visible. The parameters ($\omega, k, \theta, n$) used to plot the resonance curves are listed in Table 1. The corresponding spectral locations are pointed at with the white arrows in Figures 4 and 7.
 (Multimedia view with an accompanying movie starting at t = 0 $\wpe^{-1}$ and ending at 3000 $\wpe^{-1}$. Multimedia view)
  } \label{Fig2}
\end{figure}

\begin{figure}
\centering
\includegraphics[width=0.99\linewidth]{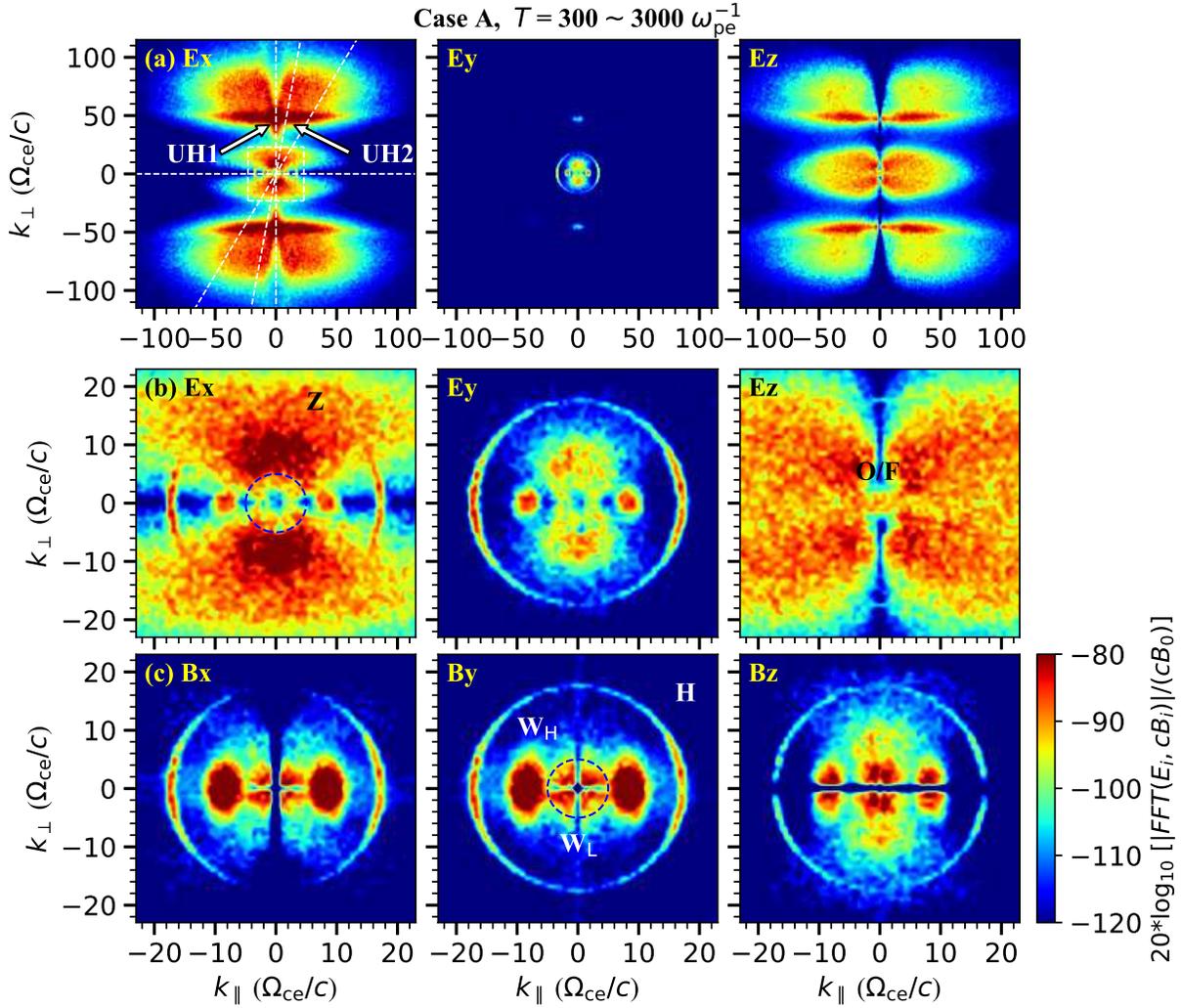}
  \caption{Wave intensity maps of the six field components in the wave-vector ($\vec k$) space for Case A. The dashed lines in panel (a) represents $\theta= 0^{\degr}$, $60^{\degr}$, $80^{\degr}$, and $90^{\degr}$ along which dispersion relations are plotted in Figures 4 and 6. Middle and lower panels are the zoom-in views of the white square in panel (a). The blue circles in panels b and c delineate the spectral regime of the $W_L$ mode. (Multimedia view with an accompanying movie starting at t = 0 $\wpe^{-1}$ and ending at 3000 $\wpe^{-1}$. Multimedia view)}
\label{Fig3}
\end{figure}

\begin{figure}
\centering
\includegraphics[width=0.99\linewidth]{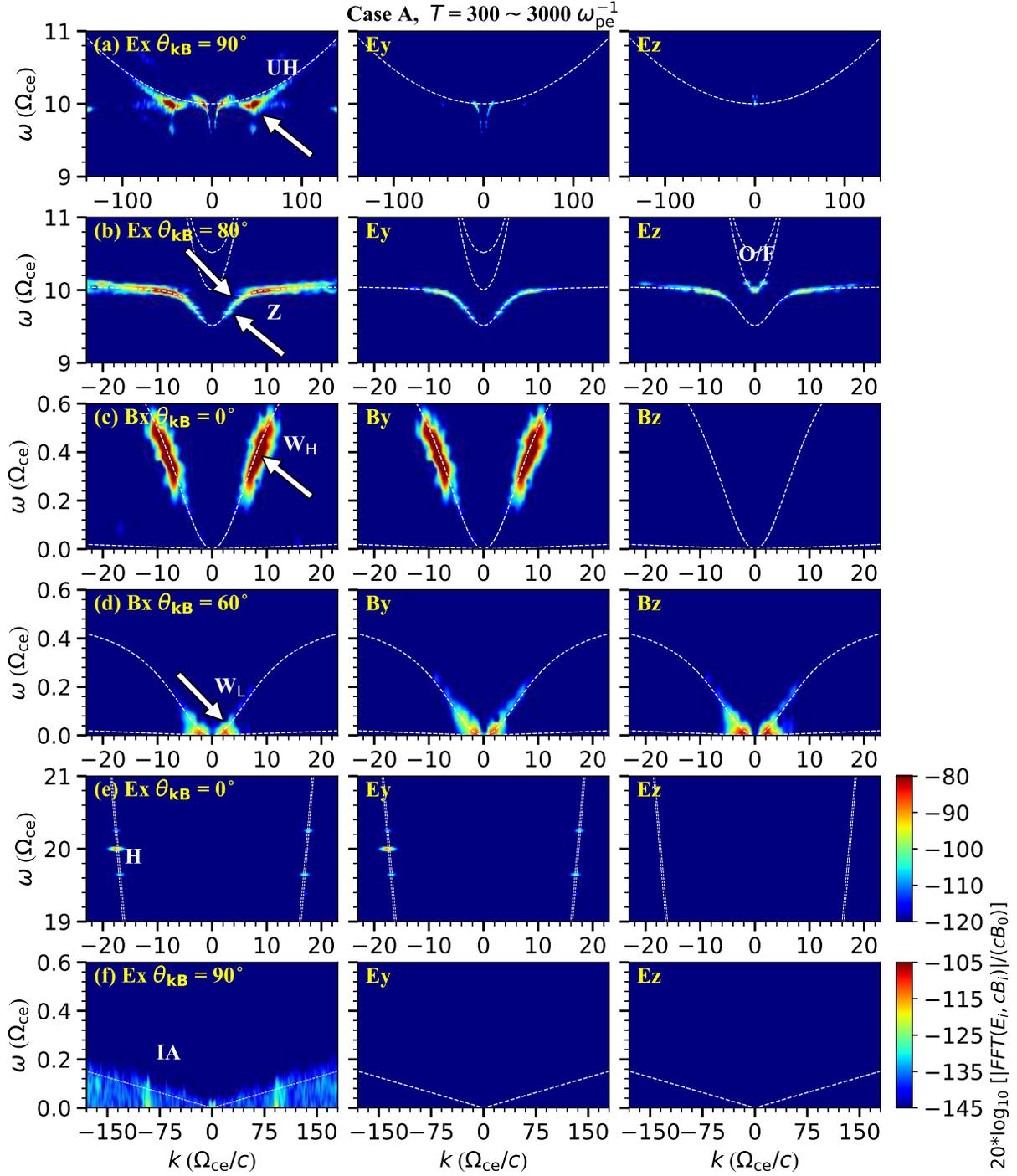}
  \caption{Dispersion diagrams of (Ex, Ey, Ez) and (Bx, By, Bz) to
  present the UH (a), Z (b), W$_\mathrm{H}$ and W$_\mathrm{L}$ (c), H (d), and IA (e) modes.
  Analytical dispersion curves of corresponding magnetoionic modes are over-plotted. The data have been processed using the Hanning windowing to alleviate the effect of spectral leackage. (Multimedia view with an accompanying movie starting from $\theta = 0^\circ$ and ending at $\theta = 90^\circ$. Multimedia view)}
\label{Fig4}
\end{figure}

\begin{figure}
\centering
\includegraphics[width=0.99\linewidth]{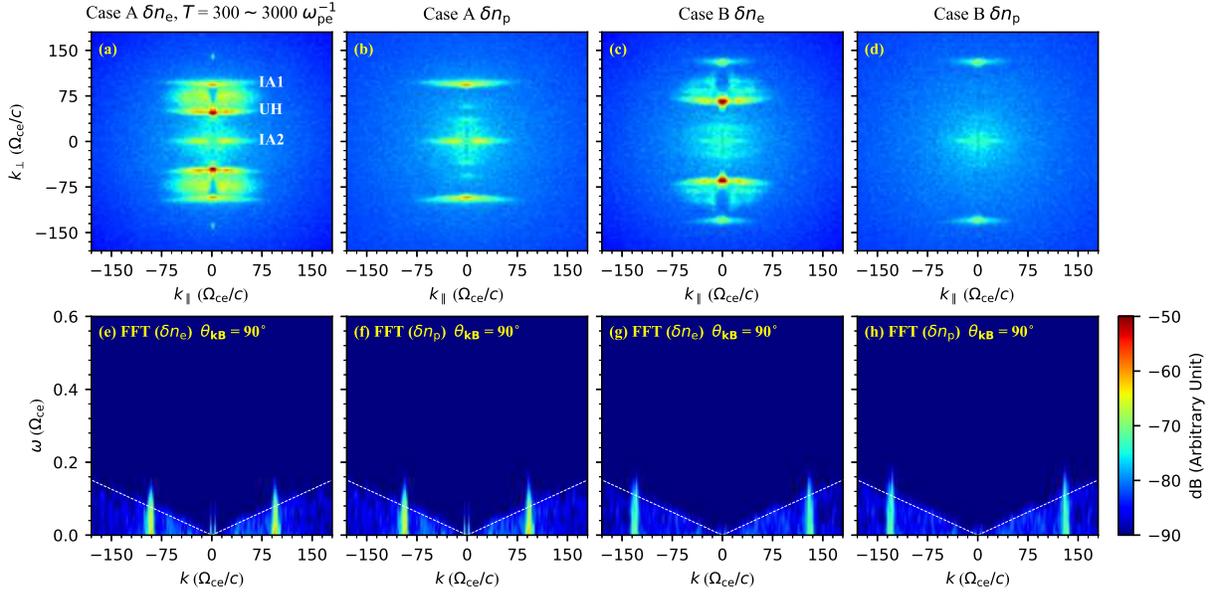}
  \caption{Upper panels: the spectral diagrams in the $\vec k$ space for density
  fluctuations of electrons ($\delta n_\mathrm{e}$) and protons ($\delta n_\mathrm{p}$);
  lower panels: the corresponding $\omega_k$ dispersion diagrams along $\theta = 90^{\degr} $
  for Cases A (right four panels) and B (left four panels). The lines represent the standard IA dispersion relation. (Multimedia view with an accompanying movie starting at t = 0 $\wpe^{-1}$ and ending at 3000 $\wpe^{-1}$. Multimedia view)}
     \label{Fig5}
\end{figure}

\begin{figure}
\centering
\includegraphics[width=0.99\linewidth]{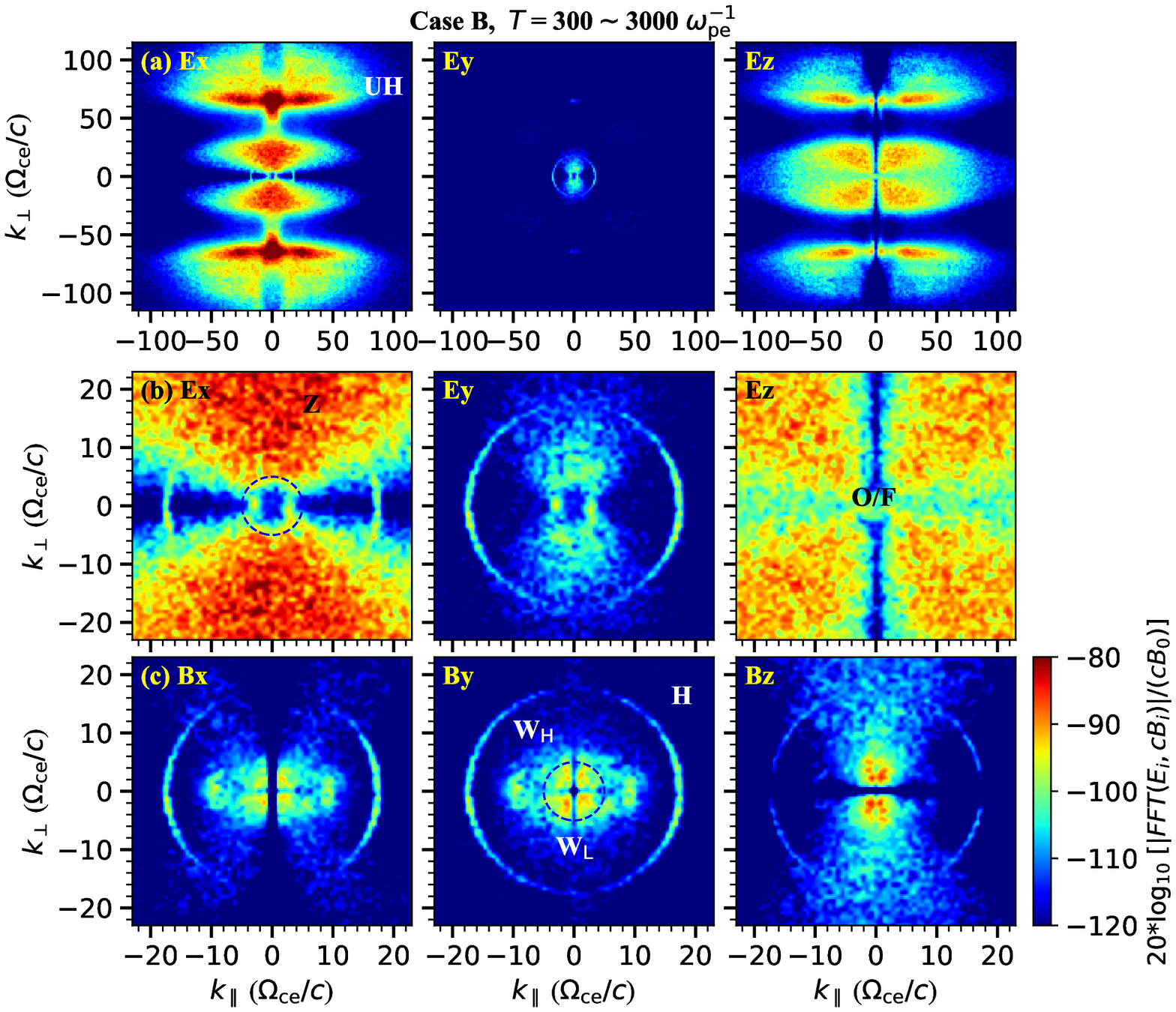}
  \caption{The same as Figure 3 yet for Case B.}
     \label{Fig6}

\end{figure}

\begin{figure}
\centering
\includegraphics[width=0.99\linewidth]{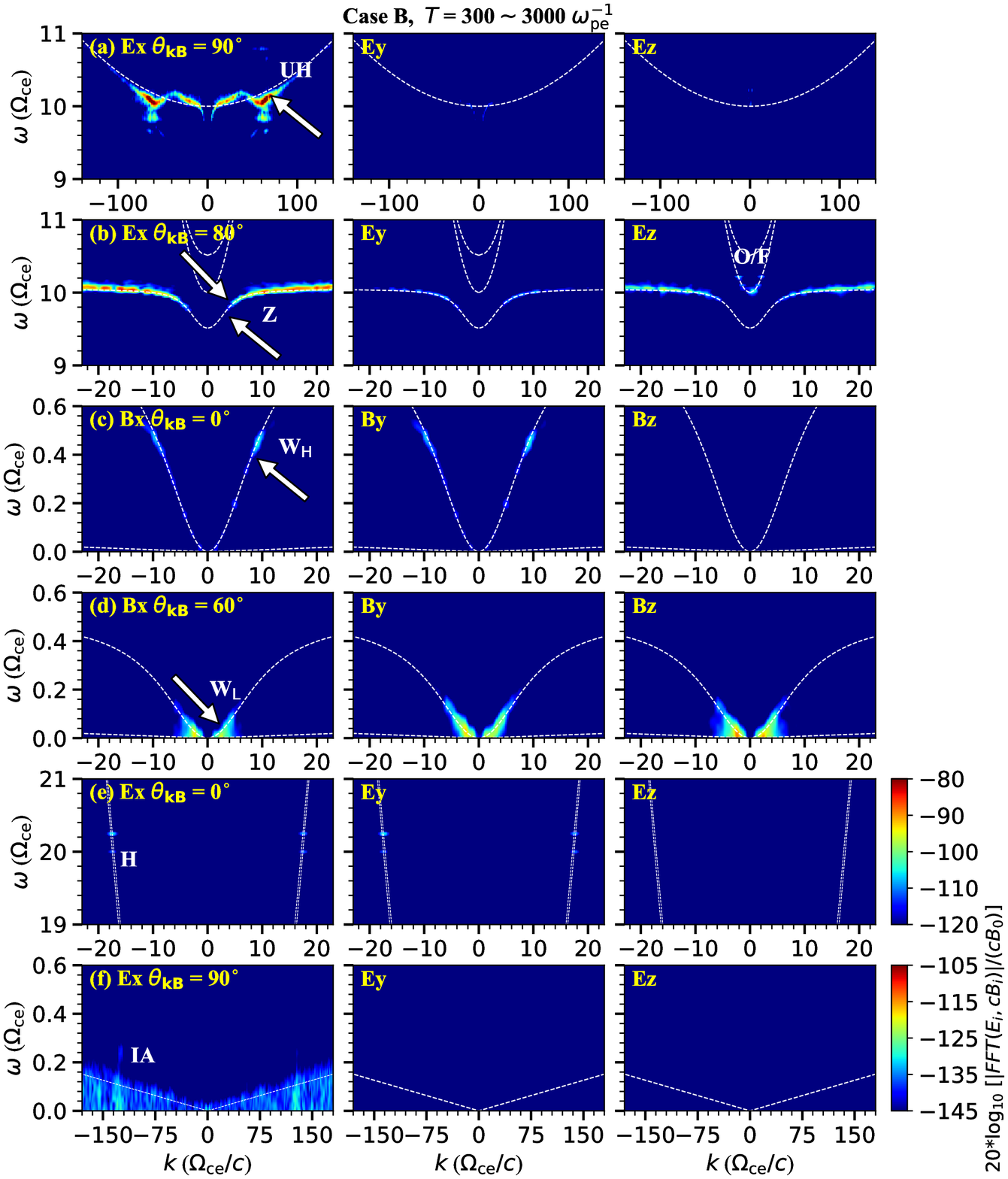}
  \caption{The same as Figure 4 yet for Case B. }
     \label{Fig7}

\end{figure}

\begin{figure}
\centering
\includegraphics[width=0.99\linewidth]{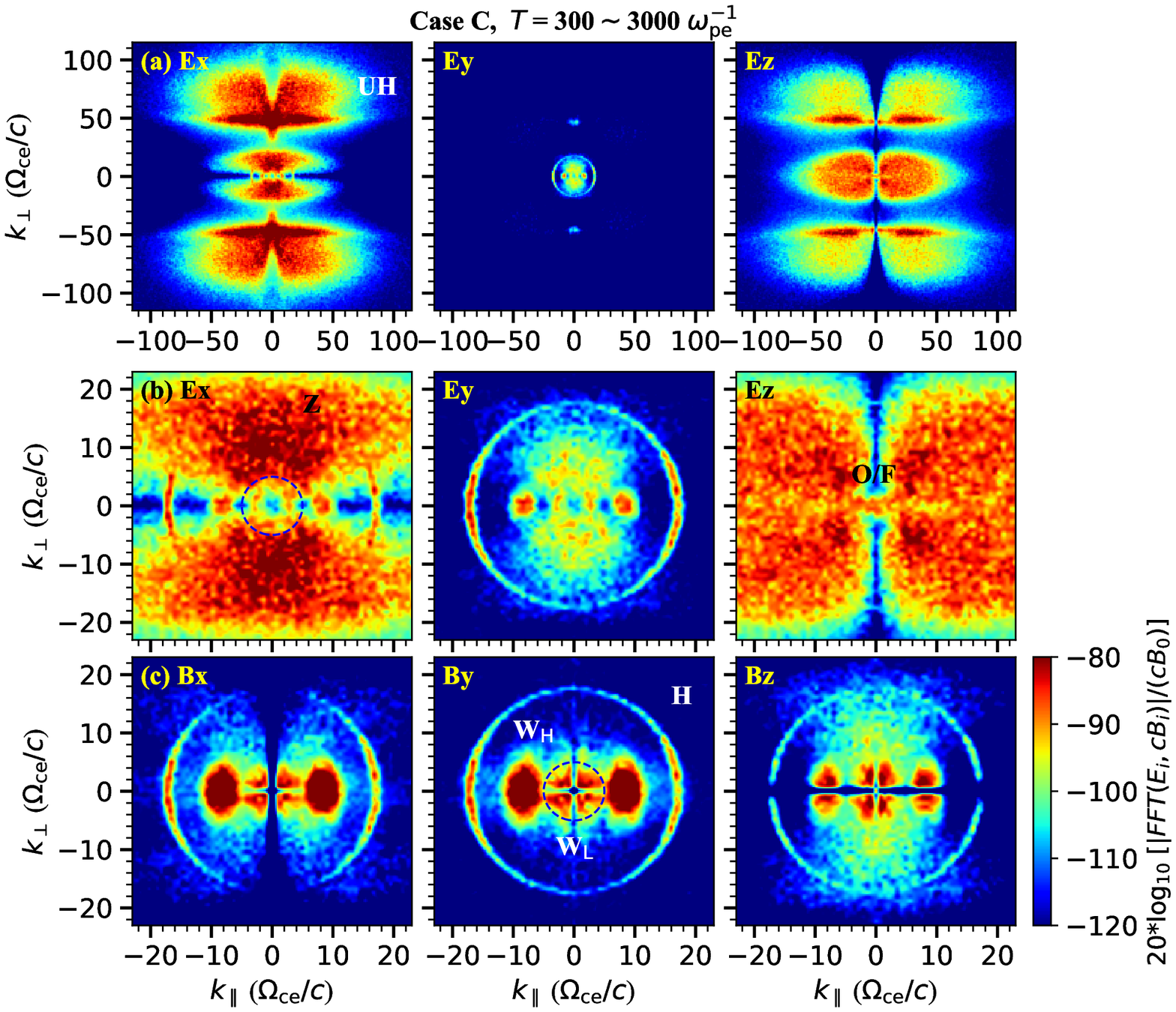}
  \caption{The same as Figure 3 yet for Case C.}
     \label{Fig8}

\end{figure}

\begin{figure}
\centering
\includegraphics[width=0.99\linewidth]{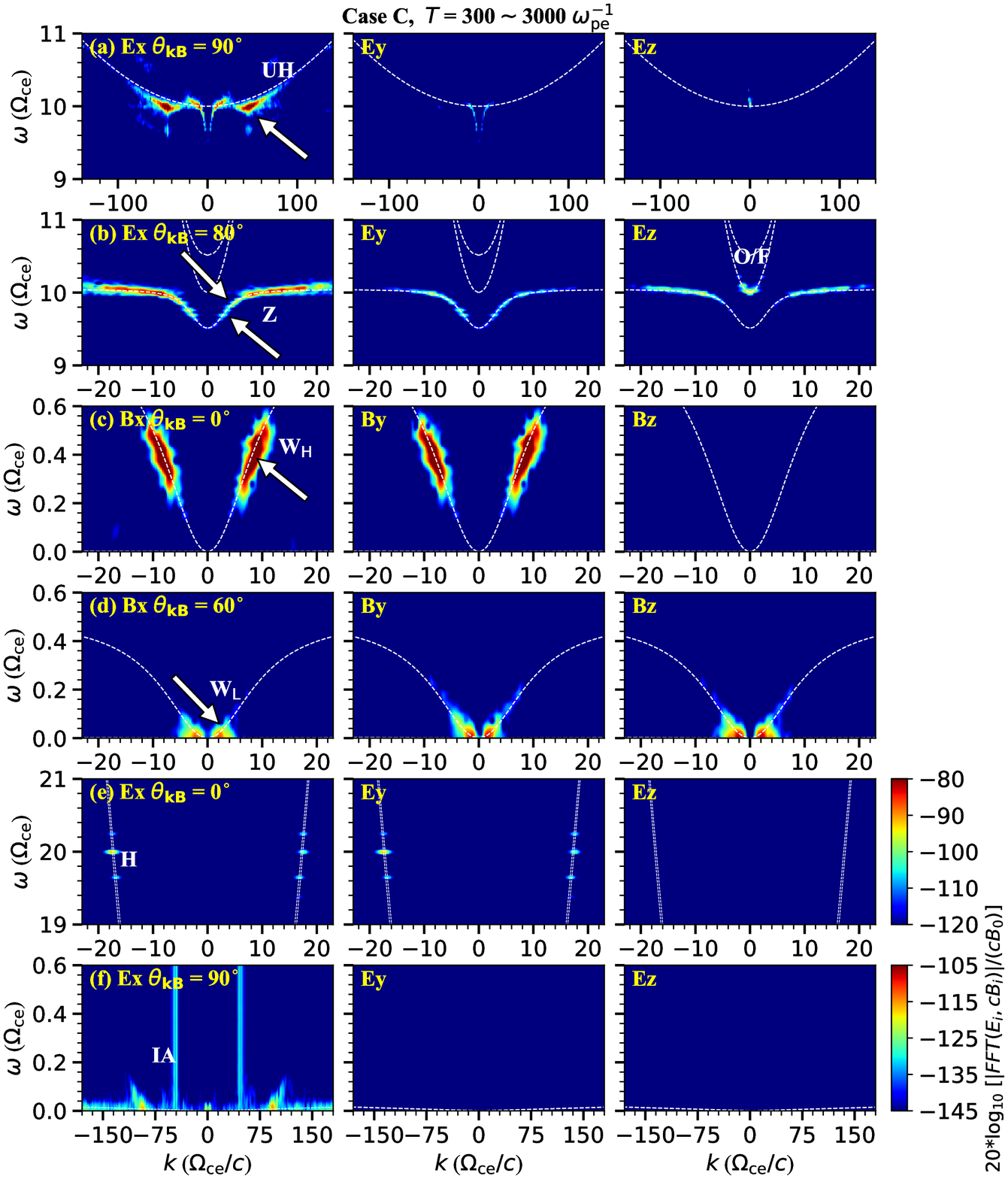}
  \caption{The same as Figure 4 yet for Case C.}
     \label{Fig9}
\end{figure}

\end{document}